\documentclass{aa} 
\usepackage{graphicx}
\usepackage{gensymb}
\usepackage[utf8]{inputenc}
\usepackage{mathpazo}
\usepackage{amsmath}
\usepackage{amssymb}
\usepackage{multirow}
\usepackage{url}
\usepackage{multicol}
\usepackage{subcaption}
\usepackage{stfloats} % For placing figures at the bottom/top of the page
\usepackage{wrapfig}
\usepackage{booktabs}
\usepackage{tabularx}
\usepackage{rotating}
\usepackage{threeparttable}
\usepackage{adjustbox}
\usepackage{array}
\usepackage{float}

\usepackage{txfonts}
\usepackage[hidelinks]{hyperref}
\hypersetup{
    colorlinks=true,
    allcolors=blue,}

\begin{document}

  \title{
Super-resolving {\it Herschel} - a deep learning
based deconvolution and denoising technique
  }

  \subtitle{I. First application at 500 $\mu$m in the COSMOS field}

  \author{Dennis Koopmans \inst{1,2}\thanks{\email{D.Koopmans@sron.nl}}
  \and Lingyu Wang \inst{1, 2}
  \and Berta Margalef-Bentabol \inst{1}
  \and Antonio La Marca \inst{1, 2}
  \and Matthieu Bethermin \inst{3}
  \and Laura Bisigello \inst{4}
    \and Zhen-Kai Gao \inst{5, 6}
    \and Claudia del P. Lagos \inst{7, 8, 9}
    \and Lynge Lauritsen \inst{10}
      \and Stephen Serjeant \inst{10}
  \and F.F.S. van der Tak \inst{1, 2}
  \and Wei-Hao Wang \inst{5}
  }

  \institute{SRON Netherlands Institute for Space Research, Landleven 12, 9747 AD Groningen, The Netherlands
  \and
  Kapteyn Astronomical Institute, University of Groningen, Postbus 800, 9700 AV Groningen, The Netherlands
  \and
  Université de Strasbourg, CNRS, Observatoire astronomique de
Strasbourg, UMR 7550, 67000 Strasbourg, France
\and
INAF-Osservatorio Astronomico di Padova, Via dell’Osservatorio
5, 35122 Padova, Italy
  \and Academia Sinica Institute of Astronomy and Astrophysics (ASIAA), No. 1, Section 4, Roosevelt Road, Taipei 10617, Taiwan
  \and
  Graduate Institute of Astronomy, National Central University, 300 Zhongda Road, Zhongli, Taoyuan 32001, Taiwan
  \and
International Centre for Radio Astronomy Research (ICRAR), M468, University of Western Australia, 35 Stirling Hwy, Crawley, WA 6009, Australia
  \and
ARC Centre of Excellence for All Sky Astrophysics in 3 Dimensions (ASTRO 3D)
\and
  Cosmic Dawn Center (DAWN), Denmark
  \and School of Physical Sciences, Faculty of Science, Technology, Engineering \& Mathematics, The Open University, Walton Hall, Kents Hill, Milton Keynes MK7 6AA, UK
  }

  \date{Received -; accepted -}

% \abstract{}{}{}{}{} 
% 5 {} token are mandatory
 
 \abstract
  {}
  {Dusty star-forming galaxies (DSFG) dominate the far-infrared (FIR) and sub-millimetre (sub-mm) number counts, but single-dish surveys at these wavelengths suffer from poor angular resolution, making identifications of multi-wavelength counterparts difficult. 
  Prior-driven deblending techniques require extensive fine-tuning and struggle to process large fields. 
  This work aims to develop a fast and reliable deep-learning based deconvolution and denoising super-resolution (SR) technique. 
  }
  {We employ a transformer neural network  
  to improve the resolution of the \textit{Herschel}/SPIRE 500 $\mu$m observations by a factor of 4.5, with input comprised of \textit{Spitzer}/MIPS 24$\mu$m and \textit{Herschel}/SPIRE 250, 350, 500$\mu$m images. The network was trained on simulations from SIDES and SHARK.  
  To mimic realistic observations, we injected instrumental noise into the input simulated images, while keeping the target images noise-free to enhance the de-noising capabilities of our method. We evaluated the performance of our method on simulated test sets and real JCMT/SCUBA-2 450 $\mu$m observations in the COSMOS field which have superior resolution compared to {\it Herschel}. 
  }
  {Our SR method achieves an inference time of  $\sim1s/deg^2$ on consumer-grade GPUs, much faster than traditional deblending techniques. Using the simulation test sets, we show that fluxes of the extracted sources from the super-resolved image are accurate to within 5\% for sources with an intrinsic flux $\gtrsim$ 8 mJy, which is a substantial improvement compared to blind extraction on the native images. Astrometric error is low, at $\lesssim$ 1$\arcsec$ compared to the 12$\arcsec$ pixel scale. In terms of reliability and completeness, $\gtrsim$ 90\% of the extracted sources brighter than $\sim3$ mJy are reliable and more than 90\% of the input sources with intrinsic fluxes $\gtrsim5$ mJy are recovered. 
  When applied to the real 500 $\mu$m observations, the fluxes of the extracted sources from the super-resolved map agree well with the SCUBA-2 measured fluxes (after converting the 450 $\mu$m fluxes to 500 $\mu$m using a correction factor of 0.84) for sources above $\sim10$ mJy. Our 500 $\mu$m number counts are also consistent with previous SCUBA-2 measurements. Thanks to its speed, our technique enables SR over hundreds of $deg^2$ without the need for fine-tuning, facilitating statistical analysis of DSFGs.
  }
  {}

  \keywords{techniques: image processing; galaxies: evolution; galaxies: high-redshift; galaxies: photometry; galaxies: statistics}

  \maketitle

\section{Introduction}

The angular resolution of astronomical imaging data is of key importance to  scientific exploitation, particularly for single-dish far-infrared (FIR) and sub-millimetre (sub-mm) surveys which typically have a factor of 10 or more worse resolution compared to surveys at other wavelengths (e.g. \citealt{Skelton2014, Herschel-SPIRE-Griffin2010, GOODS-ALMASurvey}). The poor resolution  gives rise to several challenges, e.g., large positional uncertainty, flux boosting and source blending or source multiplicity \citep{2014MNRAS.444.2870W, valiante2016}. Consequently, it can be extremely difficult to derive accurate and precise flux densities of individual dusty star-forming galaxies (DSFG), which dominate the FIR/sub-mm source counts, and to match them to their multi-wavelength counterparts. 
An incomplete multi-wavelength view severely limits our ability to understand the physical properties of DSFGs, such as their stellar masses, star-formation rates (SFR), active galactic nuclei (AGN) content, as well as any plans for follow-up observations. 

Current methods that tackle these limitations due to poor resolution include various deblending approaches which use prior information typically consisting of positions and fluxes of known sources detected in imaging surveys at other wavelengths with high spatial resolutions. For example, the super-deblending work of \citet{Liu_SuperDeblending_2018} and \citet{Jinetal2018} uses a prior list comprised of sources detected in the radio by the \textit{Very Large Array} (VLA) at 10 cm and 20 cm and by the Multiband Imaging Photometer for \textit{Spitzer} (MIPS) at 24 $\mu$m. One key aspect is the progressive way in which deblending is performed, fitting imaging data in decreasing order of resolution and appending new sources to the prior list that appear in the residuals due to varying depths and coverage of the surveys. Similarly, the probabilistic de-blender XID+ \citep{XIDplus_2017, Pearson2017XID, 2018A&A...615A.146P, 2021A&A...648A...8W} uses a Bayesian approach to derive the FIR/sub-mm fluxes of the sources in the prior list. It can also be applied progressively, deblending imaging data observed with \textit{Spitzer} (MIPS) at 24 $\mu$m, \textit{Herschel}
 Photodetector Array Camera and Spectrometer (PACS) at 100 and 160 $\mu$m, 
Spectral and Photometric Image Receiver (SPIRE) at 250, 350 and 500 $\mu$m, and \textit{James Clerk Maxwell Telescope} (JCMT) SCUBA-2 at 450 and 850 $\mu$m. One of the key advantages of XID+ is its ability to take into account correlations between sources by constructing joint posterior distributions. 

However, these deblending methods have significant limitations which restrict their applications particularly in large fields. First, they are very time-consuming. Both the selection of the prior sources and the deblending involve long computation times, particularly for progressive deblending. Second, the selection of the prior sources is of key importance to the success of deblending. As such detailed multi-wavelength information are needed to reliably predict if a source is likely to emit significantly in the FIR/sub-mm. In practice, even in premier extragalactic fields such as the GOODS and COSMOS fields which have extensive multi-wavelength data, the quality, coverage and depth of the data inevitably vary across the field, leading to variations in the reliability of the selection of the prior sources. Over the next few years, both the quantity and quality of high resolution multi-wavelength imaging will increase dramatically. For example, \textit{Euclid} will survey one third of the sky in the visible and near-IR bands \citep{EuclidInstrument}, complemented by ground-based surveys such as the Rubin observatory \citep{RubinObservatory}. This avalanche of deep data means that on the one hand we should be able to find counter-parts for the vast majority of DSFGs detected in the FIR/sub-mm, but on the other hand it will be even more challenging to use traditional deblending methods as the number of potential counter-parts will increase sharply. The long computing time also means that it is impractical to apply traditional methods to large fields beyond several deg$^2$. 

To tackle these challenges, one could exploit deep-learning (DL) based techniques which are able to learn and perform the same task but with significantly faster inference times due to their highly parallelisable nature. \citet{2024A&A...688A..20W}, presented an improved version of XID+ using the updated COSMOS2020 catalogue \citep{cosmos2020cat}, where the spectral energy distribution (SED) modelling step, crucial for prior selection, is replaced by a DL based emulator, significantly reducing computation times. \citet{Lauritsen_SR_2021} introduced a DL based super-resolution (SR) method employing an auto-encoder, a deep convolutional neural network (CNN), to predict the \textit{JCMT}/SCUBA-2 450 $\mu$m source fluxes. Their input to the network consists of the same-field observations in the three \textit{Herschel}/SPIRE bands at their native resolutions, with the PSF full width half maximum (FWHM) equal to 18.1$\arcsec$, 24.9$\arcsec$ and 36.6$\arcsec$ \citep{2010A&A...518L...5N}, respectively. The output is a super-resolved 450 $\mu$m map with the PSF FWHM set to $7.9$$\arcsec$. The network was trained on simulated SPIRE images generated from the Empirical Galaxy Generator (EGG, \citealt{2017A&A...602A..96S}) as well as observational data. The observational super-resolving target comprise of the same-field SCUBA-2 image at 450 $\mu m$, which has a PSF FWHM of 7.9$\arcsec$. Predicting the 450 $\mu$m source fluxes allows for direct comparison with the available SCUBA-2 450 $\mu $m data in the COSMOS field. Their proof of concept is a promising way forward, especially considering the less than one minute per deg$^2$ inference time and requiring no fine-tuning of the selection of source priors, making it an ideal approach to in large-area surveys. 

In this paper, we build on the SR method demonstrated in \citet{Lauritsen_SR_2021} to super-resolve the SPIRE 500 $\mu$m observations. In Sect. \ref{sec:data}, we introduce the simulated and observational data. We supplement the input priors consisting of the SPIRE imaging with the MIPS 24 $\mu$m to obtain more sparsity in the correlations at FIR/sub-mm wavelengths. We train our model solely on simulated images generated by the Simulated InfraRed Extragalactic Sky (SIDES; \citealt{2017A&A...607A..89B}) and SHARK \citep{SHARK} simulation codes. We evaluate the performance of our method through comparisons with a simulation test set and the real SCUBA-2 observations at 450 $\mu m$ of the COSMOS field. In Sect. \ref{sect:Methods}, we present a deep transformer neural network to deconvolve and denoise {\it Herschel} imaging of the COSMOS field at 500 $\mu$m with a target PSF FWHM of 7.9$\arcsec$, equal to the SCUBA-2 resolution, effectively improving the resolution by a factor of 4.5. Transformer networks \citep{Vaswani-et-al, ViT2020} are able to achieve better results in classification and regression tasks. This is due to their ability to model dependencies over a large range of variable spatial scales and a highly scalable architecture, allowing for deeper networks. Besides a novel model architecture, we introduce an alternative loss function for higher photometric accuracy. 
This loss function replaces the aperture loss function used by \citet{Lauritsen_SR_2021} which was found to be non-differentiable in Python libraries such as \textit{Tensorflow} \citep{tensorflow2015-whitepaper} due to reliance on external libraries. In Sect. \ref{sect:Results}, we present the super-resolution performance on simulations and observations. Finally, in Sect. \ref{sect:Conclusions}, we discuss and summarise the results presented in this paper as well as future outlook towards a simultaneous multi-wavelength SR method.

%----------------------------------------------------------------
\section{Data}\label{sec:data}

We employ simulations of the FIR/sub-mm sky and real observations. The purpose of using the simulation data set is to train and test the neural network. The observational data consist of the observations of the COSMOS field that we will be using to super-resolve the 500 $\mu m$ imaging. The input data set $X$ consists of the {\it Spitzer}/MIPS 24 $\mu m$ image and the three {\it Herschel}/SPIRE images at 250, 350 and 500 $\mu$m, i.e., $X = \{x_{24\mu m}, x_{250\mu m}, x_{350\mu m}, x_{500\mu m}\}$. The output target set is the super-resolved 500 $\mu$m map $Y = \{y_{500\mu m}\}$ with a factor of 4.5 higher angular resolution compared to the input 500 $\mu$m map. This choice in resolution is motivated by our desire to compare the super-resolved 500 $\mu$m map, 
after we apply the trained network to the real observations, with the SCUBA-2 450 $\mu$m observations.
The neural network $G$ is tasked to learn a mapping  $G(X) \to Y$, i.e. to apply SR on the simulation-based training data. The characteristics of the input and target imaging data such as the PSF FWHM, noise and pixel scales are listed in Table \ref{table:characteristics_combined}. All imaging data have units of Jy/beam. Below we describe how we generated the simulation data and how we obtained the observational data.

\begin{table*}[ht]
    \caption{Characteristics of the \textit{Spitzer} MIPS, \textit{Herschel} SPIRE, \textit{JCMT} SCUBA-2 and super-resolved \textit{Herschel} SPIRE data. }
    \centering
	\begin{threeparttable}
		\small 
		\adjustbox{max width=\textwidth}{
			\begin{tabular}{l|c|c|c|c|c|c|c|}
				\hline
				& \multicolumn{6}{c|}{\textbf{Instrument}} \\\hline 
				\textbf{Characteristic} & \textbf{\textit{Spitzer} MIPS} & \multicolumn{3}{c|}{\textbf{\textit{Herschel} SPIRE}} & \textbf{Super-Resolved \textit{Herschel} SPIRE} & \textbf{\textit{JCMT} SCUBA-2} \\ \hline
				Wavelength & $24\mu m$ & $250\mu m$ & $350\mu m$ & $500\mu m$ & $500\mu m$ & $450 \mu m$ \\
				PSF FWHM\tablefootmark{a, b} & 5.7'' & 18.1'' & 24.9'' & 36.6'' & 7.9'' & 7.9'' \\
				Confusion noise ($\sigma_{\text{conf}}$)\tablefootmark{a, b, d} & 18 $\mu$Jy & 5.8 $\pm$ 0.3 mJy & 6.3 $\pm$ 0.4 mJy & 6.8 $\pm$ 0.4 mJy & $\sim$ 0.6 mJy & 0.6 mJy \\
				Instrumental noise ($\sigma_{\text{inst}}$)\tablefootmark{c, d} & 14$\mu Jy$ & 2mJy & 2mJy & 2mJy & 0 mJy &0.59 - 13.3 mJy \\
				Pixel scale (pre-interpolation)\tablefootmark{c} & 1.2'' & 6'' & 8.33'' & 12'' & 1'' & 1'' \\
				Pixel scale (post-interpolation) & 1'' & 1'' & 1'' & 1'' & 1'' & 1'' \\ Estimated background\tablefootmark{e} & -15 $\mu Jy$ & -2.5mJy & -3.4 mJy & -4 mJy & - & -\\
				\hline
			\end{tabular}
		}
		\begin{tablenotes}\scriptsize
			\item[a] \textit{Herschel} values from \textit{\citet{2010A&A...518L...5N}.}
			\item[b] \textit{Spitzer} value from \textit{\citet{2009ApJ...703..222L}.}
			\item[c] Values from the respective data products in the COSMOS field. $\sigma_{inst}$ is approximated from the error maps.
            \item[d] \textit{JCMT} SCUBA-2 values from \textit{\citet{2024Gao}.}
            \item[e] Median values of the estimated backgrounds from XID$+$ \citep{2024A&A...688A..20W}.
		\end{tablenotes}
	\end{threeparttable}
 \label{table:characteristics_combined}
\end{table*}

\begin{figure*}
\centering
   \includegraphics[width=16.5cm]{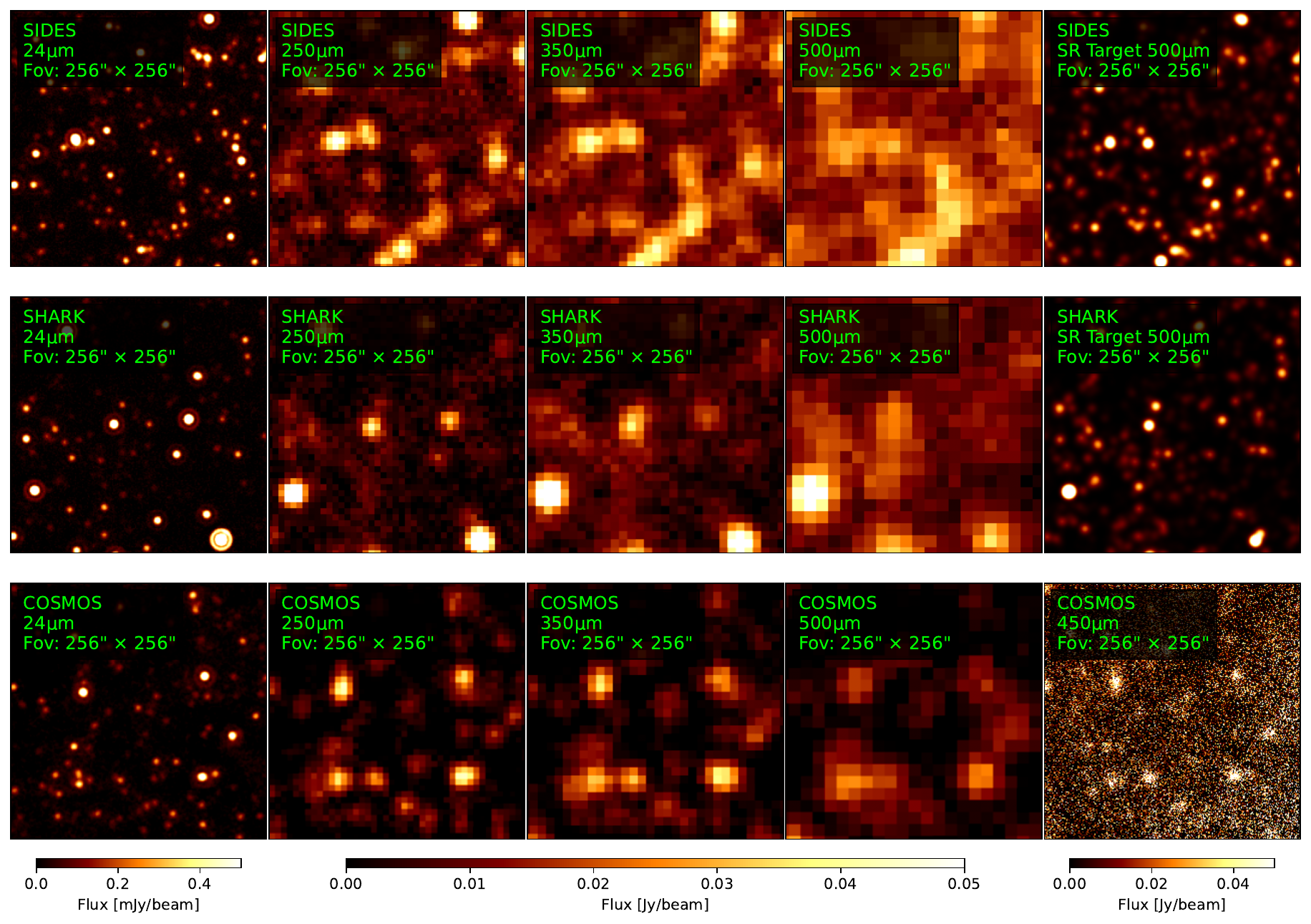}
     \caption{Example image cutouts (256$\arcsec$ $\times$ 256$\arcsec$) at MIPS 24 $\mu$m, SPIRE 250, 350 and 500 $\mu$m, and SCUBA-2 450 $\mu$m. The top two rows show simulated cutouts from SIDES and SHARK. The bottom row shows cutouts of real observations of COSMOS.  The first four columns represents the input to the network. The final column represents either the super-resolved target image at 500 $\mu$m in the case of simulations, or the SCUBA-2 450 $\mu$m image in the case of real observations.}
     \label{fig:dataset_samples.png}
\end{figure*}

\subsection{FIR/sub-mm simulations}

The performance of large, complex deep neural networks heavily relies on the quality and quantity of the training data. Especially with the additional 24 $\mu$m input image, we require a significant amount of data to train the network to learn a mapping between the input and the output. Therefore, two simulation codes (SIDES and SHARK), tailored for long-wavelength observations, are used. This allows us to include a diverse range of source flux distributions and clustering. Using the mock catalogues, generated by the simulation codes, we constructed maps using the map maker included in the Python version of SIDES, pySIDES. For all SPIRE bands, we used Gaussian PSFs with corresponding FWHMs listed in Table \ref{table:characteristics_combined}. For the MIPS 24 $\mu$m data, we used the PSF also used by the EGG, which has a Gaussian core with the FWHM listed in the same table and an outer ring. From experiments, we found that using pure Gaussian PSFs for \textit{Spitzer} data could lead to a 10\% larger error in 500 $\mu$m flux. 

The input maps $X$ are generated at their native pixel scales corresponding to the real observations, as listed in Table \ref{table:characteristics_combined}, while the super-resolved maps $Y$ are generated at a pixel scale of 1$\arcsec$. Subsequently, all input and target maps are fed into a pre-processing pipeline with the following steps: 
\begin{enumerate}
\item Add Gaussian noise (to simulate instrument noise) $\mathcal{N}(0, \sigma_{inst})$ to the input maps with the respective $\sigma_{inst}$ in Table \ref{table:characteristics_combined}. 
\item Interpolate the input maps onto a pixel scale of 1$\arcsec$. 
\item Cut the maps into smaller images of 256$\arcsec$ $\times$ 256$\arcsec$, with the requirement that the image cutouts across the input and target data maps cover the same region on the sky. 
\end{enumerate}
The size of our cutouts is around one third of the size used in \citet{Lauritsen_SR_2021}. This is because a neural network typically scales with the pixel dimensions which can result in overfitting, long training times, as well as high video memory (VRAM) usage. Finally, we use a 80\%-10\%-10\% training-validation-test split and the training set is randomly augmented during training through flipping and 90\degree \space rotations. In Fig. \ref{fig:dataset_samples.png}, we show a sample of the input and target data for each simulation after pre-processing. Below we highlight the key aspects of the simulation codes used to generate the mock catalogues and maps.

\subsubsection{Simulation Codes}
The first simulation code we employ is the Python version of the Simulated InfraRed Extragalactic Sky (pySIDES; \citet{2017A&A...607A..89B}). We used 60 deg$^2$ area from the simulated catalogues generated by \citet{CONCERTO2022} with the pySIDES code on the Uchuu dark matter simulation \citep{Uchuu2021} spanning 117 deg$^2$ and a redshift range of $0 < z < 7$. This dark matter halo catalogue contains information such as the position, mass and redshift $z$ of the haloes and is used to ensure realistic clustering through abundance matching between halos and galaxies. The mock galaxy catalogue is constructed by drawing galaxies from stellar mass functions, represented by a double Schechter function \citep{Baldry2012-schechter}, with evolution of the parameters taken into account. Some fractions of the galaxies are assigned to be star-forming, depending on their stellar mass $M_{star}$ and $z$. SIDES assumes that only star-forming galaxies emit in the FIR/sub-mm, which are further split into main sequence (MS) and starburst (SB) galaxies, with redshift-dependent probability. The average SFR $\left<SFR\right>_{MS}$ is computed as a function of $M_{star}$ and $z$, based on the observed MS. The SFRs for the galaxies are drawn from a distribution centered at 0.87$\times \left<SFR\right>_{MS}$ and $5.33 \times \left<SFR\right>_{MS}$ for the MS and SB galaxies respectively, with an upper limit set to $1000 M_{\odot} /yr$. 
Finally, SEDs are assigned based on galaxy type, total IR luminosity $L_{IR}$ and mean intensity of the radiation field $\left<U \right>$. These are used to derive flux densities through convolution with the desired filters. $L_{IR}$ is derived from the Kennicut law \citep{KennicutLaw} relating SFR and $L_{IR}$. $\left<U \right>$ is strongly correlated with dust temperature $T_{dust}$ and determines the location of the peak in the SED.   

The second simulation employed is SHARK, which is a semi-analytical model of galaxy formation and evolution \citep{SHARK} including physical processes such as gas accretion, shock heating and radiative cooling of gas, star formation and stellar feedback, growth of AGN (via merging and accretion) and AGN feedback. All galaxies in SHARK are assumed to be composed of a disc and a bulge. To generate SEDs for simulated galaxies, two packages PROSPECT \citep{2020MNRAS.495..905R} and VIPERFISH\footnote{\url{https://github.com/asgr/Viperfish}} are used. The former combines stellar synthesis libraries with a dust attenuation model \citep{2000charlot} and dust re-emission model \citep{2014Dale}. The latter extracts star-formation and metallicity histories and then generates the SEDs in the target filters. The SHARK catalogue spans more than 107 deg$^2$ in area, from which we use 60 deg$^2$, with a redshift range of $0 < z \lesssim 6$. 
\subsection{FIR/sub-mm observations}
\label{sec:FIR/sub-mm observations}

\begin{figure}
\includegraphics[width=8.5cm]{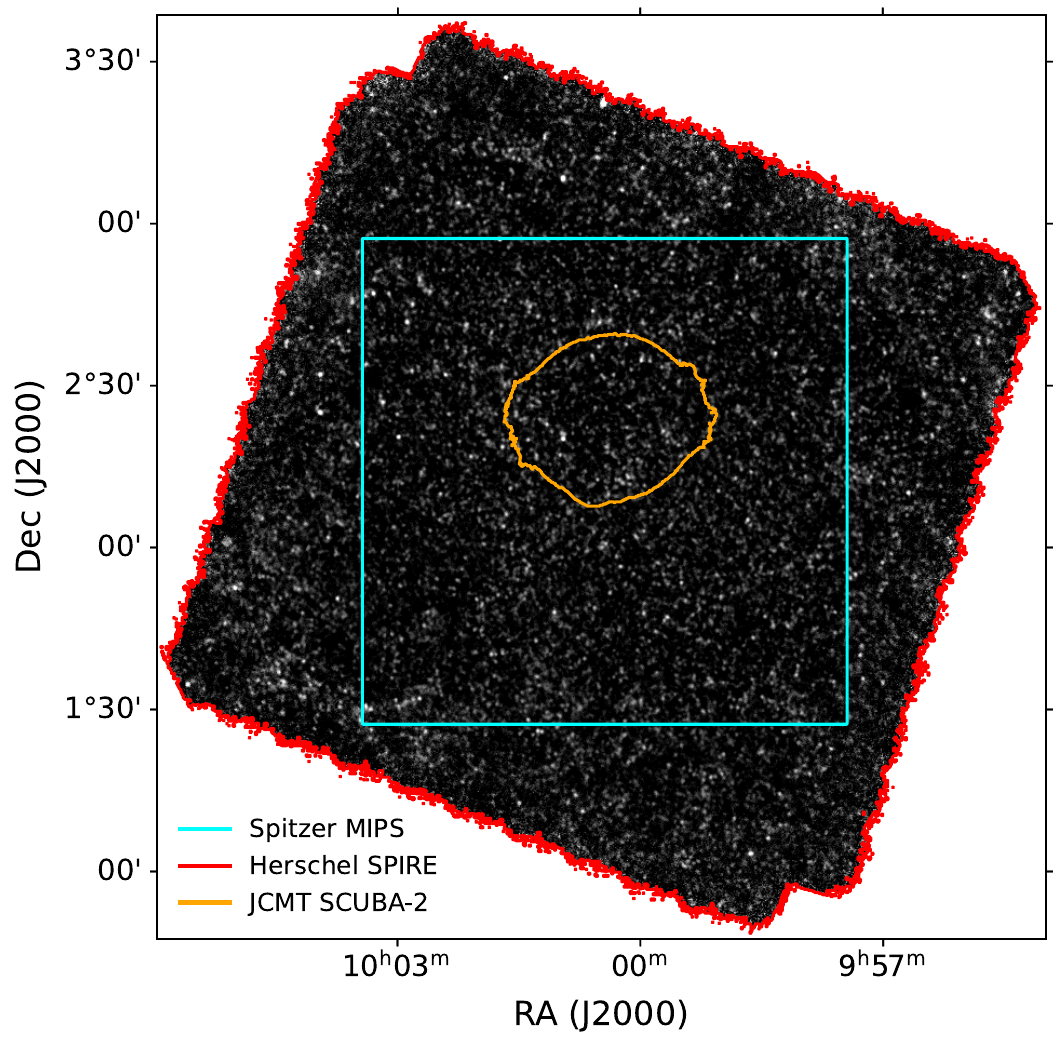}
  \caption{Contours of the MIPS, SPIRE and SCUBA-2 maps projected onto the SPIRE 500$\mu$m map. Our super-resolved 500$\mu$m map falls within the MIPS coverage.}
  \label{fig:cosmos_contour_projection}
\end{figure}

\begin{figure*}
\centering
   \includegraphics[width=16.5cm]{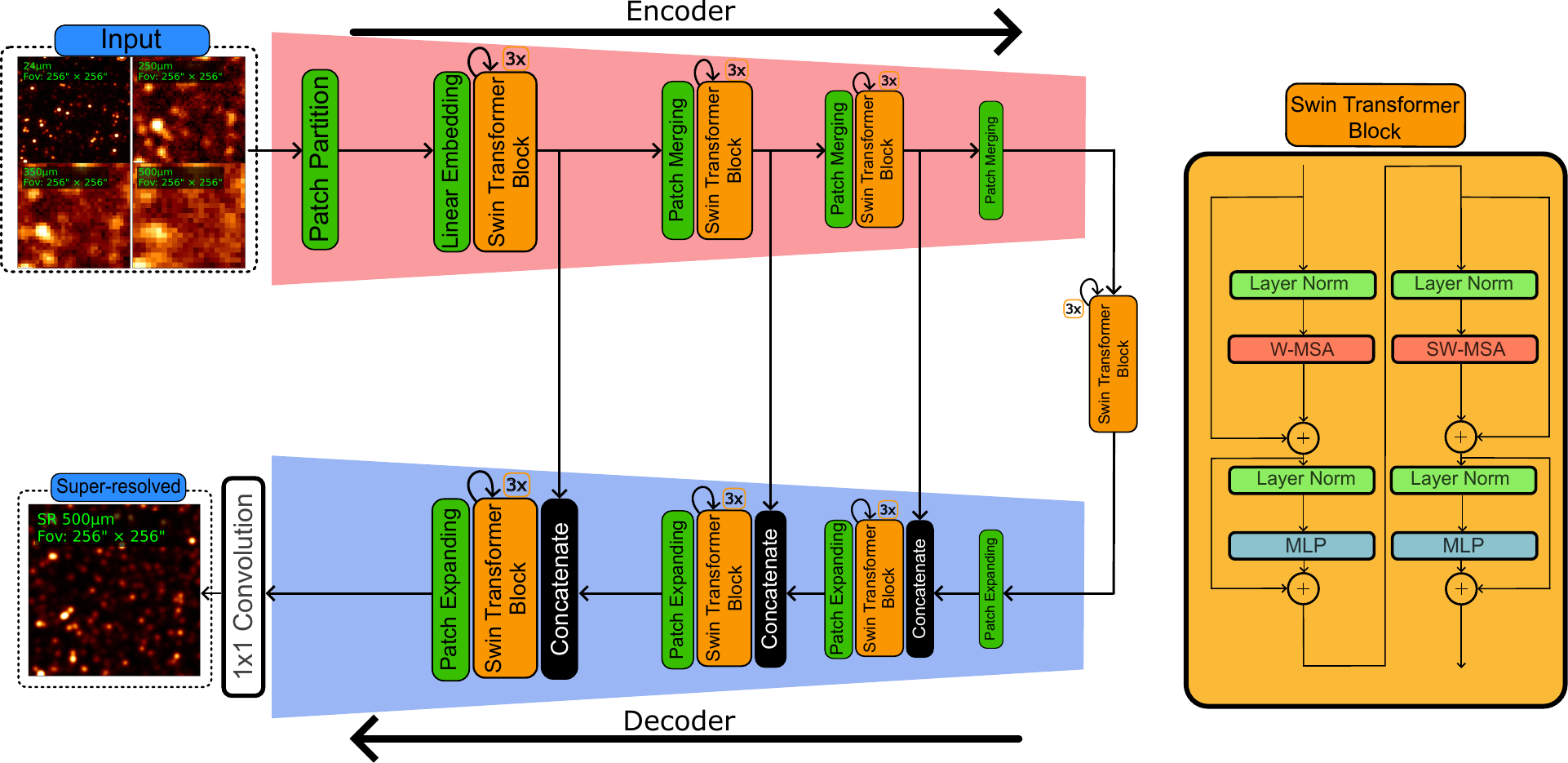}
     \caption{Architecture of the Swin-Unet network adaptation used in this paper to super-resolve the SPIRE 500 $\mu$m images. The input consists of 24, 250, 350 and 500 $\mu$m images of the same area. The output is the super-resolved 500 $\mu$m image. The network is structured as an auto-encoder, where information is compressed/decompressed by downsampling/upsampling the spatial dimensions of the features by a factor of 2, using the Patch Merging and Patch Expanding layers. The encoder and decoder is connected by skip connections (concatenations) at various dimensional levels, with a bottleneck layer at the bottom. The image is split in patches in the Patch Partition layer and embedded together with its relative position in the Linear Embedding layer. In each encoder/decoder level, the Swin Transformer Block is applied three times in consecutive order. This Swin Transformer Block is also depicted on the right, comprised of the windowed multi-head self-attention (W-MSA) part and a part where the window is shifted (SW-MSA). Finally, in our adaptation an additional $1\times1$ convolutional layer at the output is added such that we obtain a super-resolved image.}
     \label{fig:SwinUnetArch}
\end{figure*}

The Cosmic Evolution Survey (COSMOS) field \citep{2007ApJS..172....1S}, covering $\sim2$ deg$^2$, is a popular field with an extensive array of observations across many wavelengths. We use the MIPS 24 $\mu$m, SPIRE 250, 350, and 500 $\mu$m, and SCUBA-2 450 $\mu$m maps, which are pre-processed in the same way as the simulations, with the exception of adding noise. Two datasets are created, one comprised of the COSMOS field covered by MIPS and SPIRE and one over a much smaller region that has SCUBA-2 coverage, as shown in Fig. \ref{fig:cosmos_contour_projection}. 

The MIPS 24 $\mu$m we used is the GO3 map from the COSMOS-Spitzer program \citep{2009ApJ...703..222L} \footnote{https://irsa.ipac.caltech.edu/data/COSMOS/images/spitzer/mips/}, which has a relatively high resolution with a PSF FWHM of 5.7$\arcsec$. Therefore, it contains crucial information to perform SR. There is a very small chance of low-$z$ 24 $\mu$m sources appearing as extended sources \citep{Jinetal2018} (Fig. \ref{fig:extended_source_analysis}). Given that we only train on the simulated data, which treats all galaxies as point sources, this may result in sub-optimal performance of the extended sources through unwanted stretching or hallucination of a single source as multiple sources. Detailed investigation of this potential issue is beyond the scope of this paper. More relevant is that sources at high-$z$ tend to be exceedingly faint or undetected at 24 $\mu$m. It is found that $\sim$80\% of the 24 $\mu$m emission can be attributed to galaxies at $z\leq$ 2 \citep{2009ApJ...703..222L}. As sources in the SPIRE bands are more likely to be at high-$z$ \citep{Amblard2010}, the 24 $\mu$m data are likely to be more helpful for deblending galaxies at low redshifts. Nevertheless, deblending lower redshift sources has a mutual benefit: it enhances the reliability of extracting the neighbouring higher-$z$ sources.  
The complete characteristics such as the confusion noise $\sigma_{conf}$ and pixel scale of the MIPS data is shown in Table \ref{table:characteristics_combined}. We applied a multiplicative correction factor of 1.369 to account for the side-lobes of the PSF.

The SPIRE maps in COSMOS are obtained from the Herschel Multi-tiered Extragalactic Survey (HerMES) \citep{2012MNRAS.424.1614O}\footnote{https://irsa.ipac.caltech.edu/data/COSMOS/images/herschel/spire/}. These maps are, in addition to the 24 $\mu$m maps, used as input. Therefore, the complete input set can be described as $X=\{x_{24\mu m}, x_{250\mu m}, x_{350\mu m}, x_{500\mu m}\}$. The 500 $\mu$m set corresponds to the images that will be super-resolved.   The PSF FWHM at 250, 350 and 500 $\mu m$ is $18.1''$, $24.9''$ and $36.6''$ respectively, as shown in Table \ref{table:characteristics_combined}. 
Unlike the 24 $\mu$m maps, the confusion noise $\sigma_{conf}$ in the SPIRE maps is much higher than the instrumental noise $\sigma_{inst}$.

The SCUBA-2 450$\mu$m map in COSMOS has been obtained from the STUDIES-COSMOS programme \citep{2024Gao}, which achieved an rms instrument noise level of 0.59 mJy in the deepest area of the map. Within an area with a 12$\arcsec$ radius centred at the deepest point, the instrument noise can vary significantly, up to 13.3 mJy at the edge of this circular area. In comparison, the $1\sigma$ confusion noise is estimated to be around 0.65 mJy \citep{2024Gao}. The SCUBA-2 450$\mu$m observations cover an area of $\sim0.5$ deg$^2$ with only a smaller area of roughly 300 arcmin$^2$ (with instrument noise lower than $\sim7$ mJy) suitable for image comparison with the super-resolved image given the noise levels. The PSF FWHM is 7.9$\arcsec$, a factor of 4.5 smaller than the SPIRE 500 $\mu$m map. The characteristics of the SCUBA-2 map is also listed in Table \ref{table:characteristics_combined}. The SCUBA-2 data are used to evaluate the performance of our models on the observational data as they are the best option available for direct comparison. While close in wavelength, from the SIDES simulations we find a significant difference (up to 20\%) between the 450$\mu$m and 500 $\mu$m source fluxes. Therefore, a correction is required for reasonable comparisons. 

In the bottom row of Fig. \ref{fig:dataset_samples.png}, we show example image cutouts from the real observations in COSMOS. 
Given that the MIPS and SPIRE observations are not absolutely calibrated, but mean subtracted, we estimate the background map using the XID$^+$ estimated background values of the sources fitted from \citet{2024A&A...688A..20W}. This is done by projecting the values on the sky, at their respective coordinates, and interpolating using a fine gridding. Since the XID$^+$ sky area is smaller than our coverage, we set the background levels in the outer regions to the median value of the XID$^+$ background values, listed in Table \ref{table:characteristics_combined}.

\section{Methods}\label{sect:Methods}

In this section, we first present our DL-based method to super-resolve and denoise the SPIRE 500 $\mu$m images. Then we describe the loss functions and setup for training the network.

\subsection{Network Architecture}
\label{sect:Network Architecture}

The network architecture that we employed follows a U-net like structure, first introduced by \citet{UnetPaper} and used by \citet{Lauritsen_SR_2021}. It is comprised of an encoder and a decoder with skip connections between each encoder and its corresponding decoder level at the opposite side of the network using concatenations. This effectively allows for a more direct flow of gradients that is at a much smaller distance from the input of the model alleviating the vanishing gradient problem that is common with deep neural networks enabling greater network depths. 

However, unlike the previous implementation by \citet{Lauritsen_SR_2021}, our implementation does not use convolutional layers except for the output layer. The generative network, or generator $G$, of our model is a pure transformer network where the architecture closely resembles that from \citet{SwinUnet} (Swin-Unet) with each encoder/decoder level comprised of the building blocks of a Swin Transformer (Swin-ViT; \citealt{W-MSA_ViT}) which is a variant of the well-known Vision Transformer (ViT;  \citealt{ViT2020}). Transformers rely on a multi-headed self-attention (MSA) mechanism followed by dense feedforward layers (MLP). This concept of self-attention was mainly introduced by \citet{Vaswani-et-al} for natural language models such as GPT-3 \citep{GPT-3}. Lately, they are also implemented for image tasks such as classification \citep{ViT2020, W-MSA_ViT}. The Swin-Unet has been applied on biomedical segmentation tasks \citep{SwinUnet} delivering state-of-the-art accuracies. We adapted the Swin-Unet to perform SR (which is a regression task), adding a final $1\times1$ convolution to the output layer of the model.

The overall idea of MSA is to calculate pixel-to-pixel correlations of the features within an image or feature map. Calculating pixel correlations yields an attention matrix with correlation coefficients that grows quadratically with image dimension. For an input image of size $256\times256$, both GPU memory usage and computational cost increase very quickly which would be unfeasible even for the newest hardware. Swin-ViT solves this by calculating these within a much smaller window comprised of $L\times L$ patches. Nevertheless, the transformer is able to cover a much larger range compared with a CNN. The Swin ViT Block further improves this by first implementing a windowed MSA (W-MSA) Swin-ViT Block followed by a shifted window MSA (SW-MSA), where the windows are shifted, allowing for inter-window correlations.

Fig. \ref{fig:SwinUnetArch} illustrates our implementation of the network\footnote{Our implementation is available from \url{https://github.com/dennmartko/DeepSPIRE}}. We used three encoder/decoder levels and one bottleneck layer. We set the window size $L$ to 8, 4, 4, 2 and the number of attention heads to 6, 12, 12, 12 for the respective levels. The patch size was chosen to be $4\times4$, and we used three Swin Transformer Blocks encoder/decoder level with 512 MLP nodes, denoting the number of neurons in the hidden layer. Finally, in the first layer, we chose an embedding dimension of 96.

\subsection{Loss functions}

The objective of the generator is to super-resolve the SPIRE 500 $\mu m$ images and make them indistinguishable from the target images. To realise this we guide the training of the network with a loss function that should be minimised. Our loss function has two components. The first component is the log-cosh loss $\mathcal{L}_{h}$, 
\begin{equation}
	\mathcal{L}_h = \frac{1}{m} \sum^{m}_{i=1} \sum^{N_{pixels}} \log\cosh\left(G(X)^{i} - Y^{i} \right).
	\label{eq:Loss1}
\end{equation}
Here, $X$ and $Y$ are the input and target images, $G(X)$ are the corresponding super-resolved images and $m$ is the batch size. With this loss function, we aim to recover the low-frequency information (i.e. large-scale structure) in the target. This loss component involves all pixels and so can result in a good "averaged" SR result. The super-resolved sources can be identified but with their fluxes potentially over or under-estimated. It is the most important loss in the early and mid training stages. 

We introduce a second loss component to focus training on reproducing the high-frequency structure in the target images, i.e. source fluxes and profiles. \citet{Lauritsen_SR_2021} included aperture fluxes of the detected sources. However,  
we later discovered that their implementation was not differentiable due to the use of an external python package handling source detection. Our second loss component, $\mathcal{L}_{aper}$, is a differentiable implementation of their function. In this component, we compute an image mask $Y_{mask}$ for each target image $Y$. For computational efficiency, this is done once before training. Consequently, we multiply pixel-wise $Y_{mask}$ with $G(X)$ and $Y$, subtract the two results and take the sum of the absolute values as described below, 
\begin{equation}
	\mathcal{L}_{aper} = \frac{1}{m} \sum^{m}_{i=1} \sum^{N_{pixels}} Y_{mask}^{i} \left\lvert G(X)^{i} - Y^{i} \right \rvert.
	\label{eq:Loss2}
\end{equation}
The image mask $Y_{mask}$ is constructed by pre-detecting sources brighter than 2 mJy in $Y$  and then setting all pixels  within an aperture of radius 4$\arcsec$ to 1 and all other pixels to 0. The complete loss function is a linear combination of the two components,
\begin{equation}
	\mathcal{L} = \alpha \mathcal{L}_h + (1 - \alpha)\mathcal{L}_{aper}.  
	\label{eq:CompleteLoss}
\end{equation}
Here, $\alpha$ is a hyperparameter that balances the respective losses. After a small grid-search, we settled on $\alpha = 0.999$.

\begin{figure*}
    \centering
    \includegraphics[width=16.5cm]{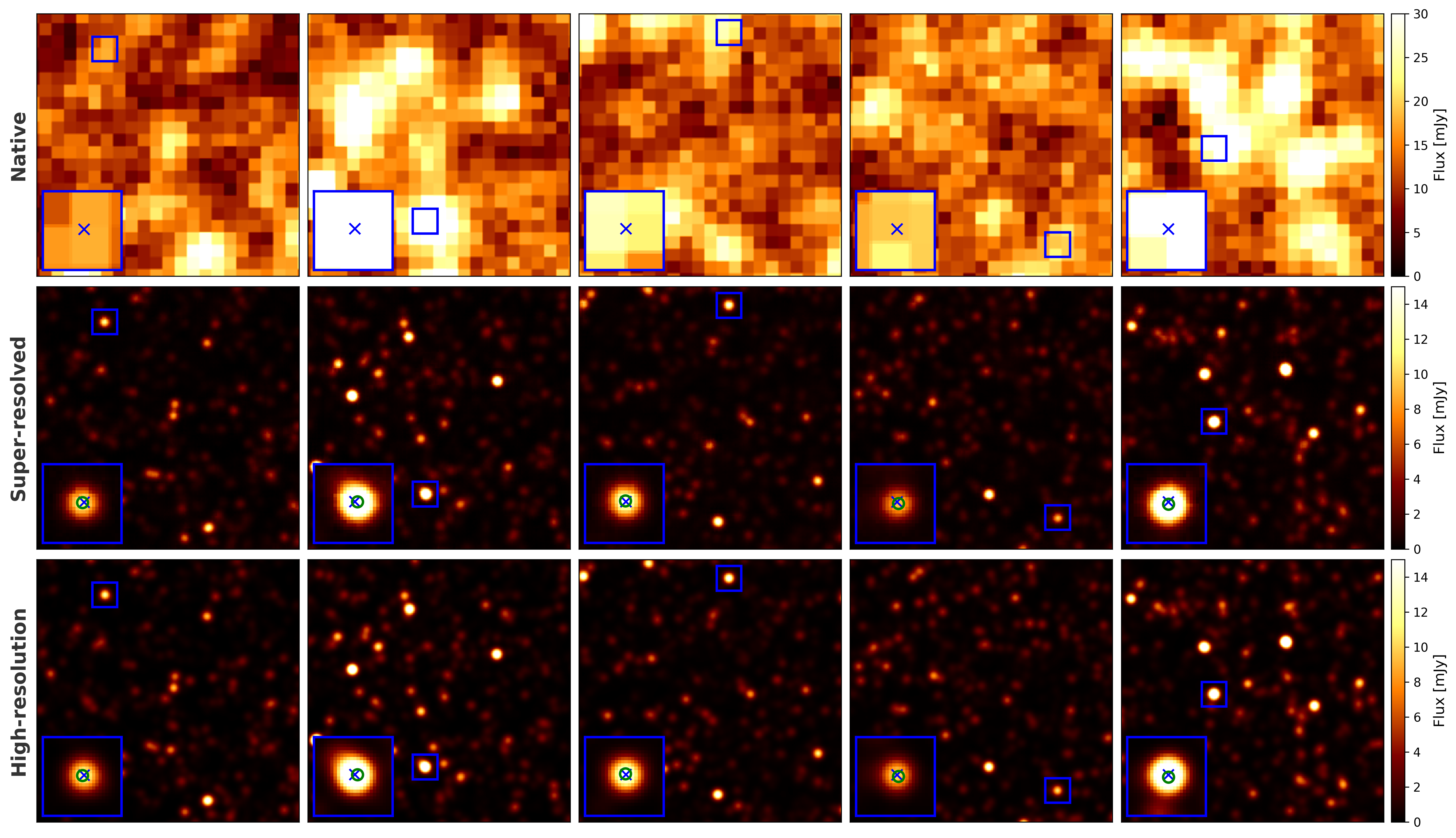}

    \caption{SR performance on the simulated images (Top: The native 500 $\mu$m images; Middle: The super-resolved images; Bottom: The target images at 7.9$\arcsec$ resolution, representing the ground truth).  
    The blue boxes highlight regions that contain at least one target source. Green circles (blue crosses) indicate sources extracted from the super-resolved (target) map.}
    \label{fig:sim_image_comparisons}
\end{figure*}

\begin{figure} % Use figure for 1-column wide figure
    \centering
    \includegraphics[width=\linewidth]{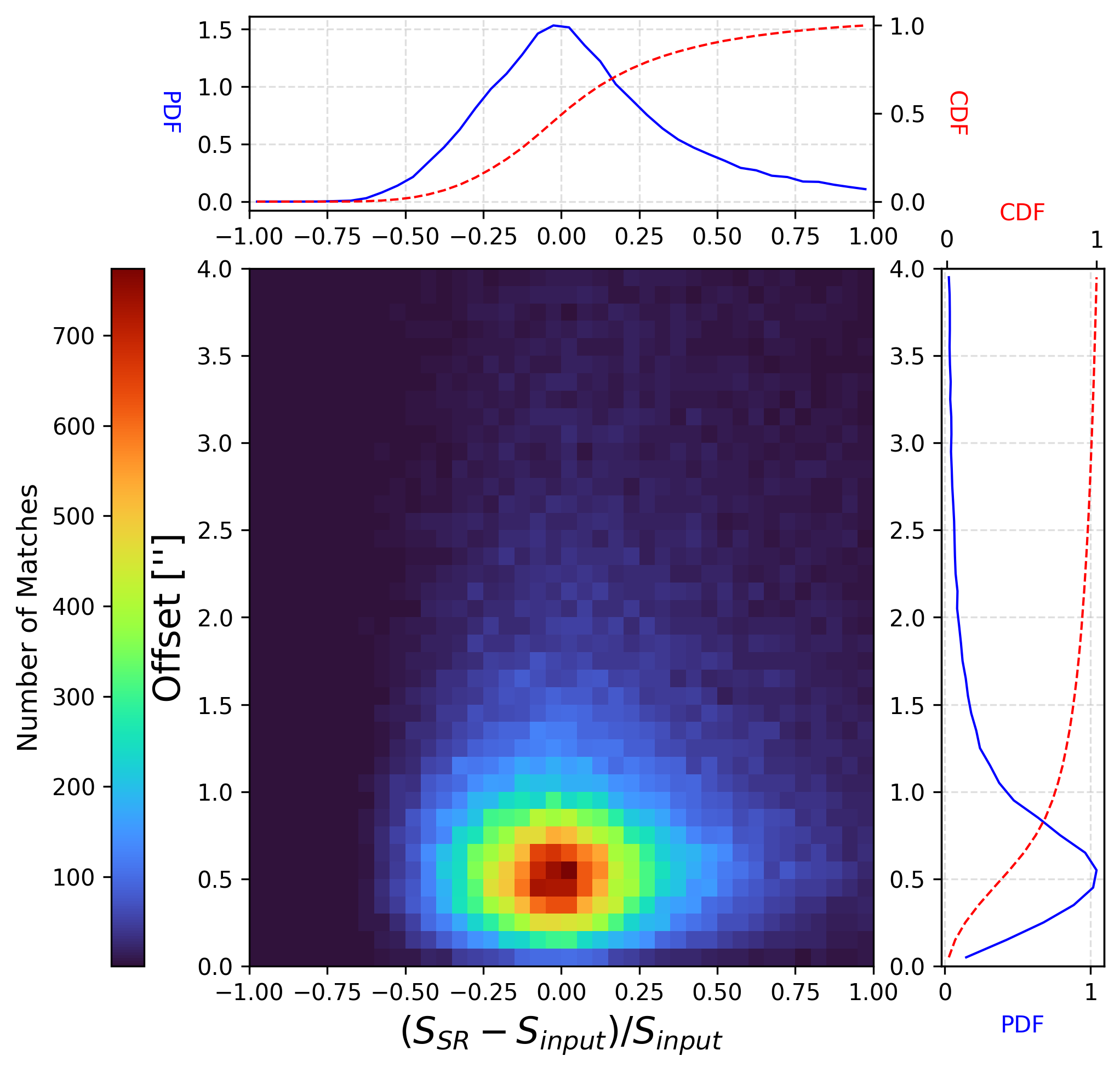} 
    \caption{Astrometric error vs. photometric error for super-resolved sources matched with the corresponding input sources using a maximum matching radius of 4$\arcsec$. 
    The marginal PDF and CDF are estimated using the counts within the selected region. 
    }
    \label{fig:offsetvflux}
\end{figure}

\begin{figure*}
  \centering\includegraphics[width=15cm]{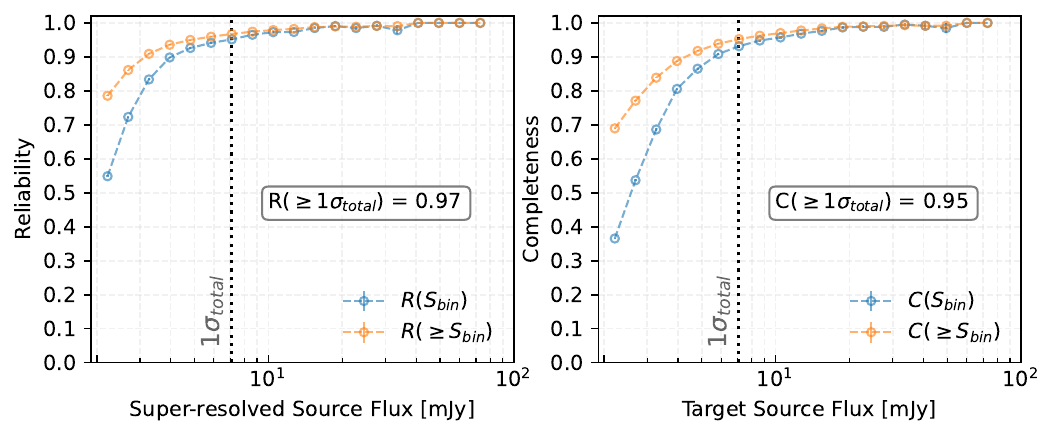}
  \caption{Left: Reliability of the super-resolved sources as a function of the detected source flux (blue: differential distribution; orange: cumulative distribution). Right: Completeness of the target sources as a function of the target source flux (blue: differential distribution; orange: cumulative distribution).
  Error bars are the $1\sigma$ uncertainties corresponding to a binomial distribution. }
  \label{fig:sim_CR}
\end{figure*}

We trained the model for up to a maximum of 5000 epochs on an NVIDIA H100 96GB GPU. During training, we used a batch size of 48 samples to calculate the gradients. The gradients are calculated using the Adam optimizer with $\beta_1 = 0.9$, $\beta_2 = 0.999$ and a learning rate $\eta$ that follows a polynomial decay scheme with power $2$, such that initially $\eta = 2\times10^{-3}$ while converging to $\eta = 1\times10^{-4}$ in $100$ epochs. This warm-up phase speeds up the training. We clip the global norm of the gradients to 7.5 to stabilize training. Further, we save the model corresponding to the lowest validation loss and we implemented early stopping with a patience of 250 epochs: if the validation loss does not improve for 250 consecutive epochs, training is terminated. After training, we found that our model required approximately 18 hours to train, with early stopping triggered at epoch 678. Finally, our model is able to super-resolve an area of $1\,\mathrm{deg}^2$ in roughly 1 second.

\section{Results}\label{sect:Results}

In this section, we present the results from our trained SR model. In Sect. \ref{sect:simresults}, we show the model performance on the simulation test set by evaluating the photometric and astrometric accuracy and provide insight into the uncertainties and limitations. In Sect. \ref{sect:obsresults}, we show the application of the model on the  observational data and compare our results with the SCUBA-2 450 $\mu$m observations.

\subsection{Performance on simulations}
\label{sect:simresults}
%------------------------
% Image comparison
%------------------------
In Fig. \ref{fig:sim_image_comparisons}, we show example comparisons of the super-resolved image cutouts with the input and target images from the simulated test set. Comparing the input 500 $\mu$m images (PSF FWHM $=36.6$") with the super-resolved images (PSF FWHM $=7.9$"), the improvement in resolution is clear. The comparison between the super-resolved images with the target images (i.e. ground-truth) demonstrates that overall our SR method is adept at recovering the intrinsic spatial distribution and fluxes of the true sources. We highlight some areas using coloured boxes, where it would be nearly impossible to extract sources from the simulated native 500 $\mu$m  images or in some cases even all SPIRE bands using blind extraction techniques. Most sources are  generated with Gaussian PSFs, although they can also occasionally appear more elongated. This is likely to be the effect of "averaging" during training and could impact on the source extraction that use PSF fitting methods causing sub-optimal flux measurements. 
Another key improvement is that while the input maps are noisy, there is no apparent presence of instrument noise in the super-resolved images. This shows that training on target images without noise can yield images with high fidelity for source extraction.

%------------------------
% Source extraction intro
%------------------------

To perform quantitative assessment, we extracted sources from our super-resolved images using a PSF fitting method from the Python package \texttt{photutils} \citep{2022ApJ...935..167A}. We use the \texttt{DAOStarFinder} function, based on the \texttt{DAOFIND} algorithm \citep{1987PASP...99..191S} to detect point sources and provide initial guesses on the source position. This function requires a minimum flux threshold and a target PSF FWHM. The algorithm searches for local peak flux densities that are above or equal to the threshold. The detected sources are then fitted using the \texttt{PSFPhotometry} function with a Gaussian PSF. We fit the central region of the PSF using a $5 \times 5$ kernel. To avoid bad fits near the image borders, we do not fit sources located within 8$\arcsec$ from the image borders.  
Determining the threshold is non-trivial, given the negligible noise  in our super-resolved images. Based on the computation time of pre-detecting sources for our loss function, we set the threshold at 2 mJy. We define sources extracted from the super-resolved and target images as super-resolved sources and target sources respectively. 

%-----------------
% Astrometry (Match radius)
%-----------------

To investigate astrometric accuracy, we derive the distribution of positional offset between the super-resolved sources and the corresponding input sources directly from the input simulated catalog. We matched the sources by selecting the input source closest in flux within a radius of 7.9$\arcsec$ (equal to the target PSF FWHM).  
Fig. \ref{fig:offsetvflux} shows the astrometric error vs. the photometric error, demonstrating no systematic correlation between the two. In addition, we plot the marginalised distribution, both the probability distribution function (PDF) and the cumulative distribution function (CDF), as a function of the relative flux difference in the top panel and positional offset in the right panel
We observe a peak in global flux underestimation at $\sim5$\%. From the CDF of the distribution of the relative flux difference, we can also infer that the $1\sigma$ confidence range in the relative difference between the super-resolved flux and the true flux is between $\sim-25$\% and $\sim25$\%. 
The positions of the majority ($\sim75$\%) of the matched super-resolved sources have sub-pixel ($\lesssim 1\arcsec$) accuracy, with a peak in the PDF at $\sim0.5$ $\arcsec$.  The high astrometric accuracy is likely due to the higher resolution of the input 24 $\mu$m image. Consequently, we  set our matching radius at 4$\arcsec$ at which we observe a near 100\% retrieval while still limiting random matches which we already started to observe near the PSF FWHM.

%-----------------------------
% Reliability and completeness
%-----------------------------
We show the differential and cumulative reliability and completeness of the super-resolved sources in Fig. \ref{fig:sim_CR}. Completeness $C$ measures the fraction of true sources that we can recover and so is defined as the ratio of target sources with a match in the super-resolved image to the total number of target sources  in the test set, as a function of the target source flux $S_{target}$. Reliability $R$ measures of the fraction of detected sources that are real and so is defined as the ratio of the super-resolved sources with a match in the target image to the total number of super-resolved sources in the test set, as a function of the super-resolved source flux $S_{SR}$. Using the widely used terms of True Positives (TP) and False Negatives (FN), and False Positives (FP), completeness and reliability can be derived as,
\begin{align}
	C(S_{\text{target, bin}}) = \frac{TP(S_{\text{target, bin}})}{TP(S_{\text{target, bin}}) + FN(S_{\text{target, bin}})},\quad C \in [0,1]
	\label{eq:Completeness} \\
	R(S_{\text{SR, bin}}) = \frac{TP(S_{\text{SR, bin}})}{TP(S_{\text{SR, bin}}) + FP(S_{\text{SR, bin}})}, \quad R \in [0,1]
	\label{eq:Reliability}
\end{align}
Our SR method shows high reliability with $\gtrsim$90\% of the super-resolved sources brighter than 4 mJy being correct and increasing to 100\% for bright sources. At the $1\sigma$ total noise ($\sigma_{total}=\sqrt{\sigma_{conf}^2 + \sigma_{inst}^2}$), both the differential and the cumulative reliability levels are very high, at around 95\%. The cumulative reliability $R(\gtrsim S_{SR, bin})$ remains $\gtrsim$80\% above our flux threshold of 2 mJy indicating that the impact of hallucinated sources is very low. The completeness shows a similar trend with almost 100\% at $\gtrsim$20 mJy. At the $1\sigma$ total noise, both the differential and cumulative completeness levels are above 90\%. The cumulative completeness $C(\gtrsim S_{target, bin})$ remains high at a level $\gtrsim90$\% for target sources brighter than 5 mJy.

\begin{figure}
    \centering
    \includegraphics[width=0.95\columnwidth]{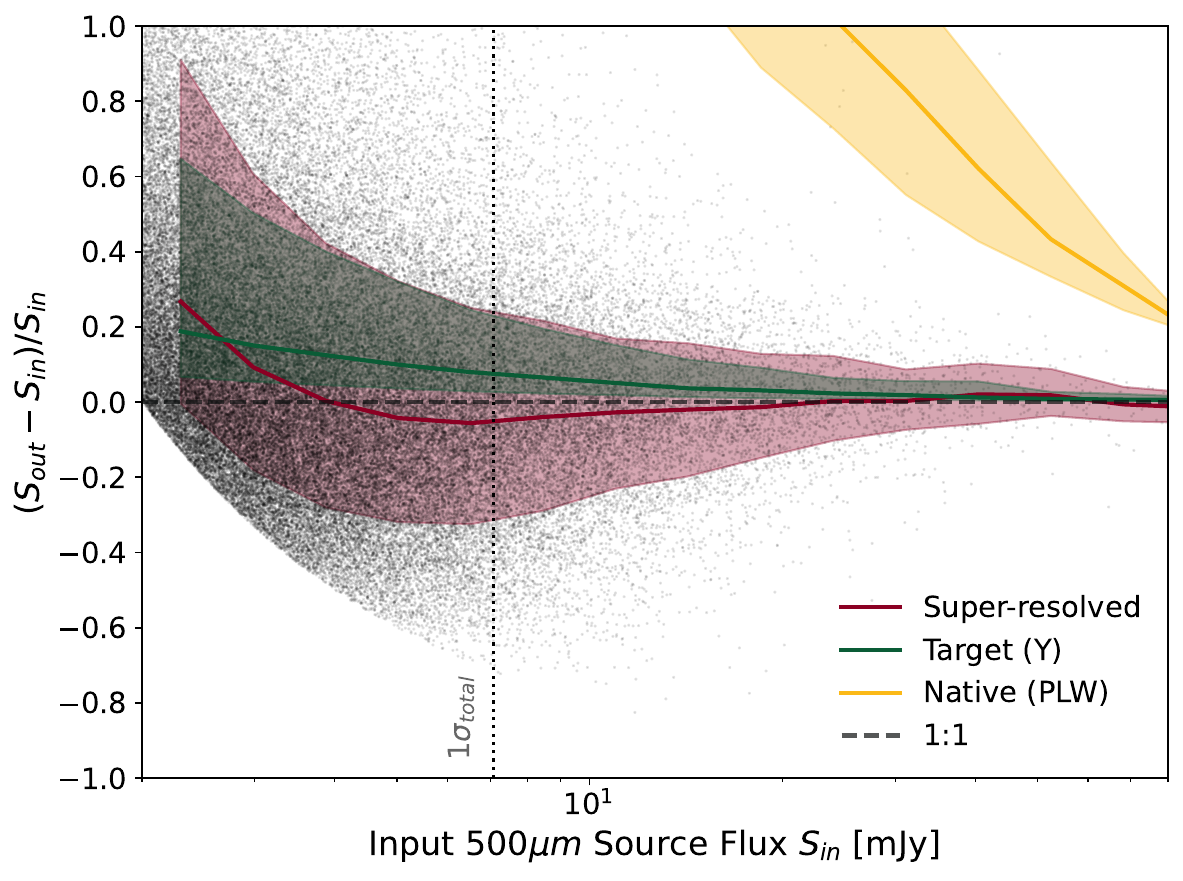}
    \caption{Fractional flux difference as a function of the input flux $S_{in}$. The measured flux $S_{out}$ corresponds to the native, target and super-resolved source fluxes. The 50th (solid line), 16th and 84th (shaded regions) percentiles are shown for each output catalogue. The dashed horizontal line indicates a 1:1 agreement. The grey points correspond to the super-resolved sources. }
    \label{fig:InputFluxReproduction}
\end{figure}

\begin{figure*}
    \centering
    \includegraphics[width=16.5cm]{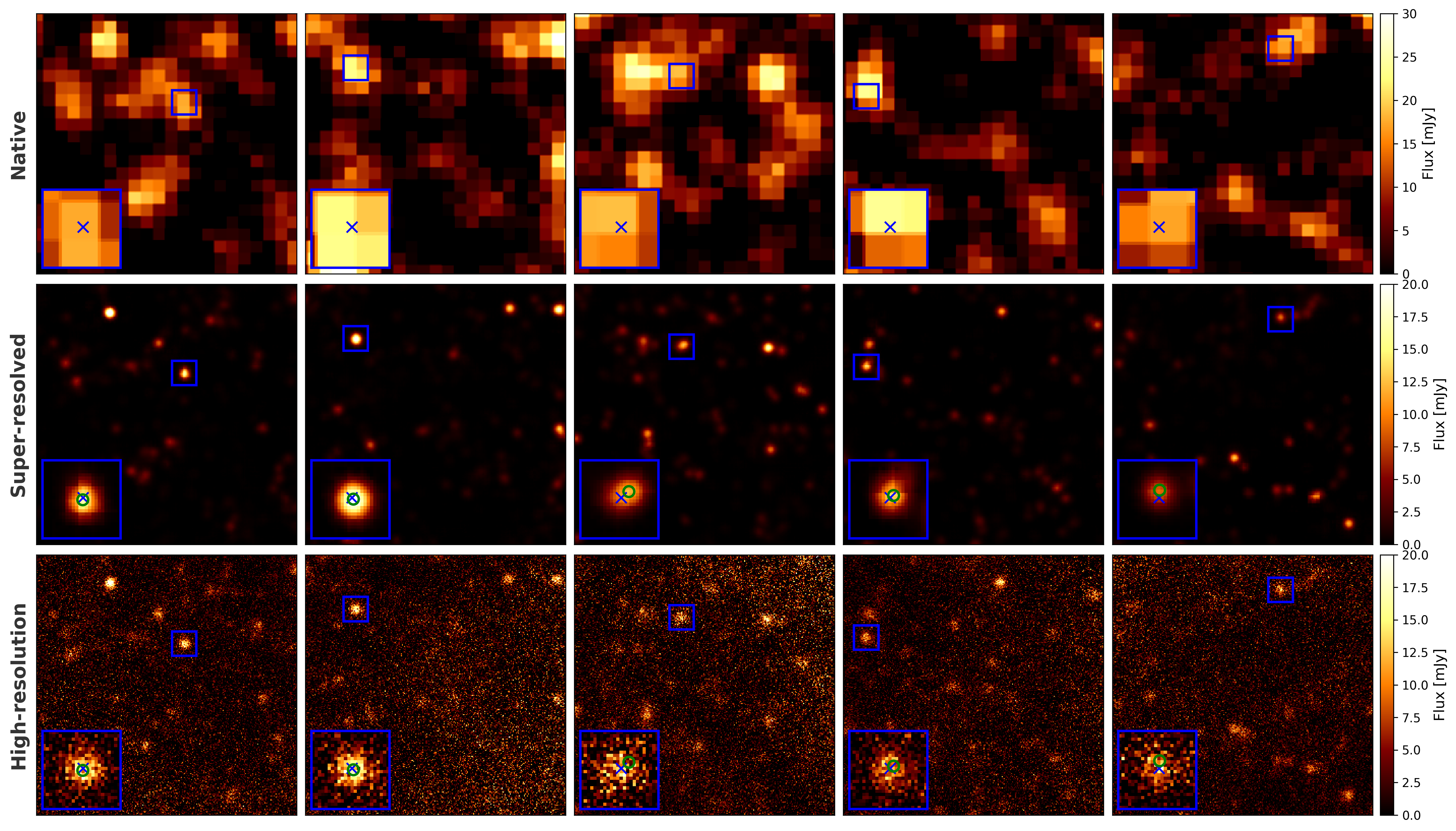}
    \caption{SR performance on the observed images (Top: The native 500 $\mu$m images; Middle: The super-resolved images; Bottom: SCUBA-2 observations converted to 500 $\mu m$, representing the approximate ground truth). The blue boxes highlight regions centred on a SCUBA-2 source. Super-resolved (SCUBA-2) sources are indicated by green circles (blue crosses).}
    \label{fig:scuba_image_comparisons}
\end{figure*}

%-------------------------------------
% Flux reproduction (Proof of Concept)
%-------------------------------------
Lastly, flux reproduction is also one of our main objectives, as recovering the input fluxes allows for more accurate SED fitting and hence better measurements of galaxy physical properties such as dust luminosities and SFRs. 
In Fig. \ref{fig:InputFluxReproduction}, we compare the fractional difference between the output flux and the true flux as a function of the true flux, with the true flux directly taken from the input simulated SIDES or SHARK catalogues. The output flux refers to the measured flux of sources extracted from the interpolated native 500 $\mu$m, target or super-resolved images, using a $5\times5$ kernel for PSF fitting. 
Fluxes of sources extracted from the native map exhibit the well-known flux boosting effect in images with poor spatial resolution.
Even at the $5\sigma$ confusion noise level ($\sim30$ mJy), the median flux boosting of fluxes measured from the native map is at around 80\%. In comparison, fluxes of sources extracted from the target images show little flux boosting effect, thanks to the much higher resolution. Generally, the median flux boosting is at around a few percent, limited to $<10$\% even at the $1\sigma$ total noise. 
Comparing the super-resolved result with the sources extracted from the native 500 $\mu$m maps, the improvement is striking. We are able to achieve much more accurate flux predictions as a function of the input fluxes even near the confusion limit. The super-resolved source fluxes follow a median underestimation of $\sim5\%$, relative to the true flux, for sources above the $1\sigma$ total noise, i.e. $\gtrsim7$ mJy. With increasing flux, the level of underestimation decreases, approaching 0\% at fluxes $\gtrsim20$ mJy. This underestimation comes from the presence of instrument noise in the input images as can be inferred from Fig. \ref{fig:InputFluxReproductionSmoothed}.
The $1\sigma$ confidence range of the relative difference between the super-resolved flux and the true flux also decreases with increasing true flux, from around 30\% at the $1\sigma$ total noise to around 5\% at the bright end above $\sim60$ mJy. As mentioned in Section \ref{sec:FIR/sub-mm observations}, most of the sources detected in Spitzer/MIPS 24 $\mu$m are low-z sources. A brief analysis on the performance as a function of the Spitzer/MIPS 24$\mu$m flux has been made in Section \ref{sec:Performance based on Spitzer MIPS}.

\subsection{Performance on observations}
\label{sect:obsresults}

%------------------------
% Image comparison
%------------------------

In this section, we show the performance of our SR method in the COSMOS field, with the input maps from the real MIPS and SPIRE observations. For the analysis of these observations, we focus on point-sources. A brief analysis on extended sources can be found in Section \ref{sec:Performance on Extended Galaxies}. 
To validate our super-resolved results, we compare with the SCUBA-2 450 $\mu$m observations (maps and extracted catalogues) from the STUDIES-COSMOS programme. Considering the low number counts, as well as our goal of measuring the number counts in the COSMOS field with high completeness, we included the extraction of sources near the image border of our super-resolved images. 

In Fig. \ref{fig:scuba_image_comparisons}, we show example image comparisons where we selected the patches containing a relatively diverse set of fluxes, number of sources and clustering. Comparing with the SPIRE 500 $\mu$m map with the native resolution, our super-resolved map clearly demonstrates a significant improvement.  
While comparison with the SCUBA-2 images is difficult due to the presence of the much higher instrumental noise, we can still to identify some counterparts that have been observed showing that our method is indeed deblending very well. 
The blue boxes highlight the regions centred on a SCUBA-2 450 $\mu$m source. The position of the source extracted from our super-resolved map agrees very well with the position of the source detected from the SCUBA-2 maps.
Moreover, our super-resolved 500 $\mu$m images show very little noise similar to what we have seen with the simulations, thanks to the de-noising capability of our SR method. In addition, the native SPIRE maps generally have lower instrument noise levels than the SCUBA-2 maps, except in the central $\sim180$ arcmin$^2$ region.
Visual inspection of the super-resolved images and the SCUBA-2 images shows that we are able to detect sources in the former in regions where their noise levels in the latter are simply too high to obtain robust detections. 

%--------------------------------
% S450 v S500SR Flux reproduction
%--------------------------------
\begin{figure*}[h]
  \centering
  \includegraphics[width=15cm]{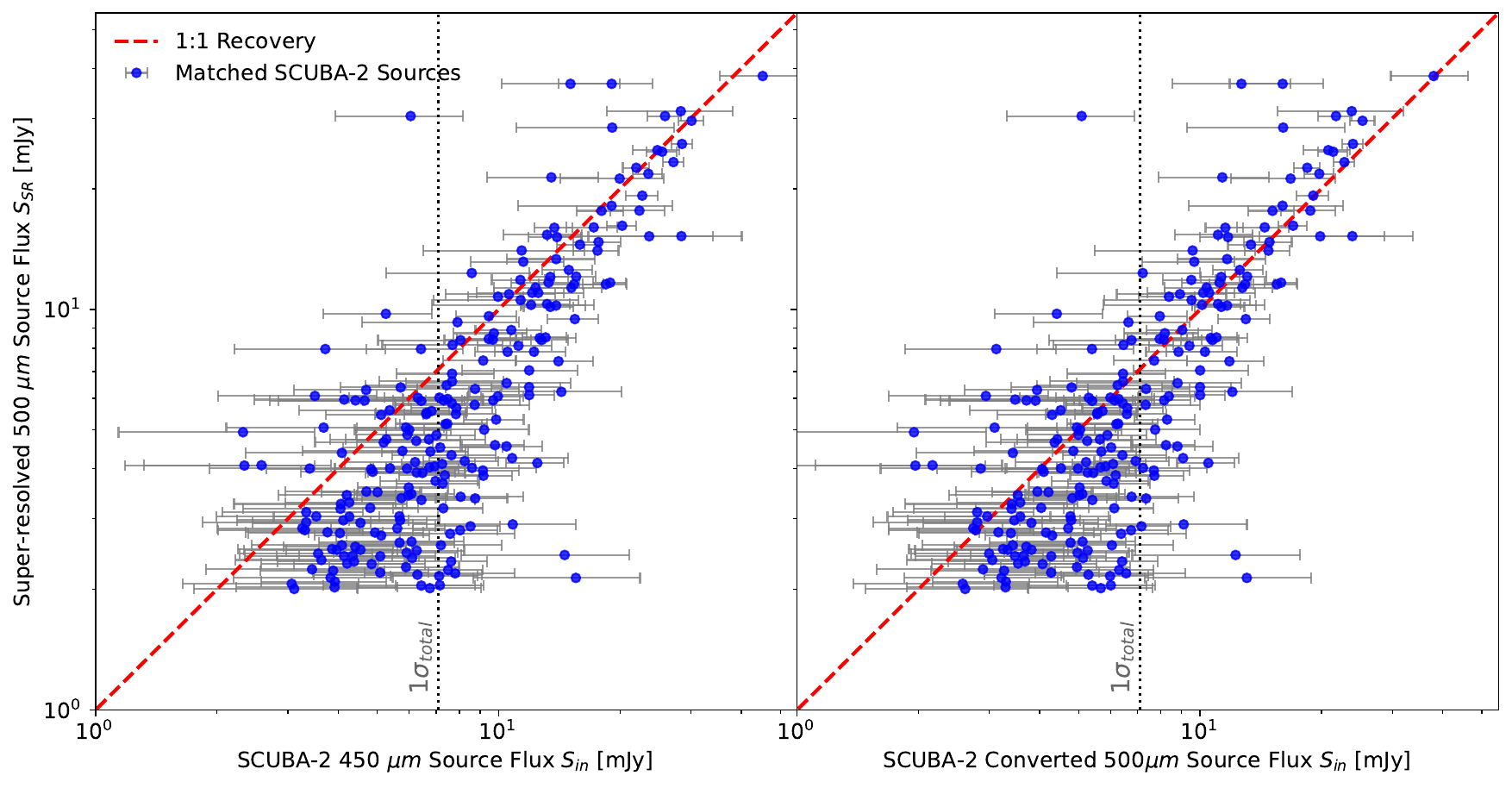}
  \caption{Our measured 500 $\mu$m flux from the super-resolved maps versus the measured SCUBA-2 flux for sources above the SCUBA-2 detection threshold of $\geq3.5\sigma$. \textbf{Left}: Comparison with the actual  SCUBA-2  450 $\mu$m fluxes. \textbf{Right}: Comparison with the 500 $\mu$m equivalent fluxes using a conversion factor from 450 $\mu$m to 500 $\mu$m. The red dashed line represents a 1:1 agreement. Error bars correspond to the measurement errors in the 450 $\mu$m fluxes and have also been converted for the 500 $\mu$m approximation in the right panel.}
  \label{fig:FluxReproduction_SCUBA}
\end{figure*}

\begin{figure*}
    \centering
    \includegraphics[width=0.85\textwidth]{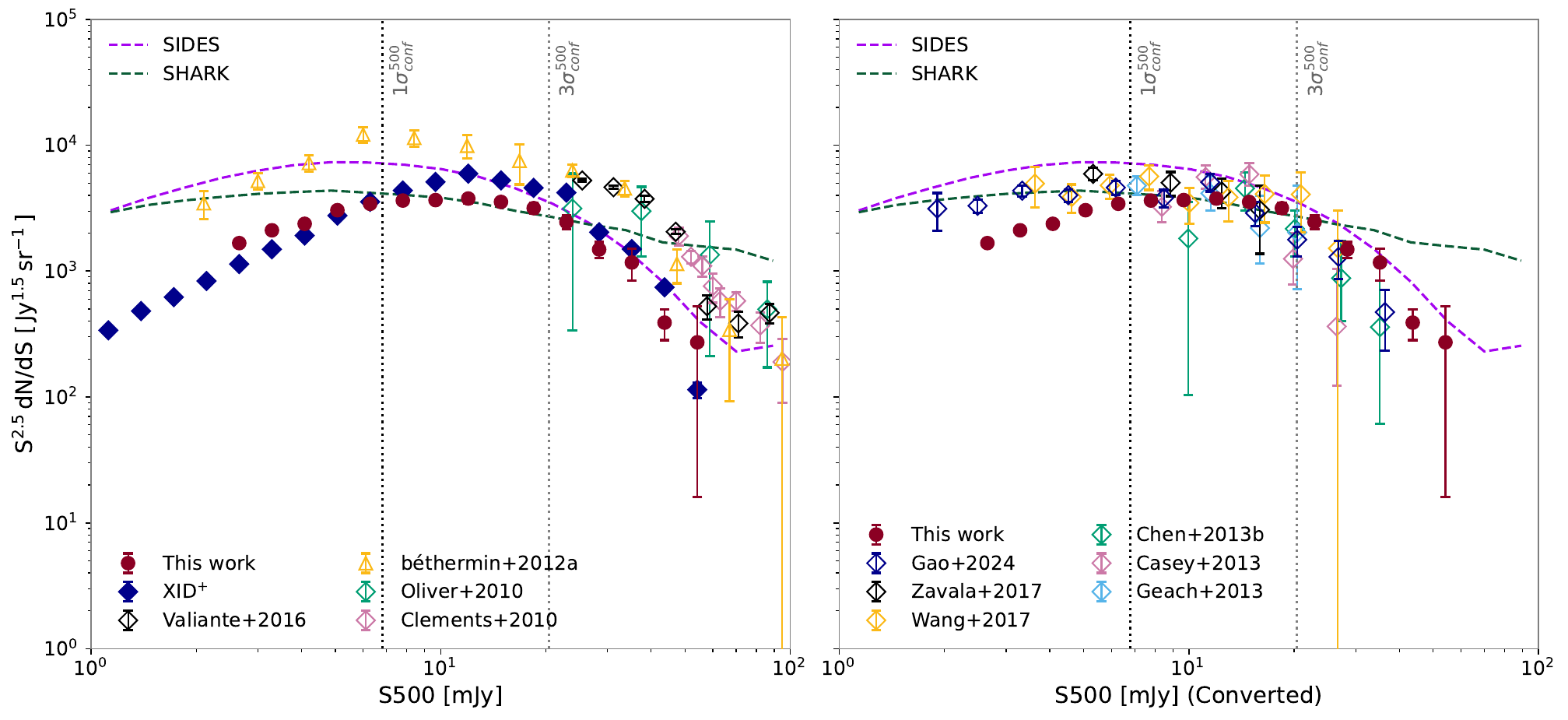}
    \caption{Comparison of the re-normalised differential number counts of our work with other counts in the literature as well as the counts produced by the simulations used in this paper (SIDES and SHARK). \textbf{Left}: Comparison of our 500 $\mu$m corrected super-resolved counts with other 500 $\mu$m counts in the literature obtained by deblending \citep{2024A&A...688A..20W}, blind extraction \citep{valiante2016, Oliver2010, Clements2010} and stacking \citep{Bethermin2012a}. \textbf{Right}: Comparison of our 500 $\mu$m corrected counts with the 450 $\mu$m counts converted to 500$\mu$m from the literature obtained by blind extraction \citep{2024Gao, zavala2017, Wang2017, Chen2013b, Casey2013, Geach2013}.}
    \label{fig:ncounts}%
\end{figure*}

Next, we conduct quantitative assessment of the level of agreement between our super-resolved fluxes and the SCUBA-2 measurements. The SCUBA-2 450 $\mu$m and our super-resolved 500 $\mu$m fluxes follow approximately a linear trend as shown in the left panel of Fig. \ref{fig:FluxReproduction_SCUBA}. 
To convert the SCUBA-2 450 $\mu$m fluxes to the SPIRE 500 $\mu$m flux, we used the $S_{450}/S_{500}$ ratio from the SIDES simulation in the flux range of $S_{450} = \left[2-100\right]$ mJy. We found a mean and median of 0.85 and 0.84, respectively. The $S_{450}/S_{500}$ distribution is multi-modal, with peaks at 0.84, 0.81 and 0.9 in decreasing order of significance. We found the median value of the $S_{450}/S_{500}$ distribution more suitable as it coincides with the highest peak. For comparison, \citet{2019WangL} used a correction factor 0.86. Using our conversion factor we display the comparison between our super-resolved source fluxes and the SCUBA-2 converted 500 $\mu$m source fluxes in the right panel of Fig. \ref{fig:FluxReproduction_SCUBA}. The mean and median ratio between the super-resolved 500  $\mu$m flux and the SCUBA-2 converted 500   $\mu$m flux is 0.95 and 0.86, respectively.
For SCUBA-2 sources above $\sim10$ mJy, the correlation is approximately 1:1 with some scatter, demonstrating reliable SR performance. 
For these relatively bright sources, the mean and median ratio between the super-resolved flux and the SCUBA-2 converted flux is 1.01 and 1.0, respectively.
For the fainter sources, while the two fluxes still agree within $1\sigma$ for most sources, there is an increased systematic overestimation of the SCUBA-2 fluxes compared to the super-resolved fluxes, which is consistent with the mild flux boosting effect seen in Fig. \ref{fig:InputFluxReproduction}.

%----------------------------------------
% XID+ vs SR and Number counts comparison
%----------------------------------------

In Fig. \ref{fig:ncounts}, we compare our corrected 500 $\mu$m number counts with those found in the literature. To estimate errors on the counts, we performed bootstrapping, randomly filling a new catalogue totalling the area of our super-resolved 500 $\mu$m catalogue, with smaller area catalogues of 0.2 deg$^2$ with replacement. This is done 10 000 times. We then computed the mean and standard deviation of the counts in each bin. Additionally, we used our super-resolved and input catalogue from simulations to compute the necessary correction factor $\mathrm{f}_c$ which corrects for reliability, completeness and flux boosting. This factor is calculated as $\mathrm{f}_c = \frac{H_{in}(S_{in})}{H_{SR}(S_{SR})}$ with $H_{in}$ and $H_{SR}$ being the histogram counts of the input and super-resolved catalogue. Here, for $H_{SR}$, we redid the source extraction and included sources near the border. For the calculation of $\mathrm{f}_c$, we used the same bins as in the calculation to estimate the count errors. We performed bootstrapping, randomly replicating the two catalogues with replacement. This is done 10 000 times and we took the mean and standard deviation of $\mathrm{f}_c$ in each bin. Consequently, the average counts in each bin have been multiplied by $\mathrm{f}_c$ and their respective errors have been propagated. In the left panel of Fig. \ref{fig:ncounts}, we can see a significant difference between the input number counts in the SIDES and SHARK simulations, in particular at the bright end. This could be due to the intrinsic differences in the construction of the two simulations, with SIDES being an empirical model and SHARK being a semi-analytic model. Our counts are closer in the overall shape to the SIDES counts at fluxes above the $1\sigma$ confusion noise. In the flux range between $\sim10$ and 30 mJy, our counts agree well with the SHARK counts. At the bright end $>30$ mJy, the SHARK counts predict significantly more sources. 
Compared to previous observational studies of the 500 $\mu$m counts in the literature obtained by blind source extraction \citep{valiante2016, Oliver2010, Clements2010}, stacking \citep{Bethermin2012a}, and prior-driven deblending \citep{2024A&A...688A..20W}, our super-resolved counts shows systematically lower abundances at almost all flux levels. The counts from blind extractions are expected to suffer from severe flux boosting which would make the bright-end of the number counts much higher than the true counts.  
In the right panel of Fig. \ref{fig:ncounts}, we compare our 500 $\mu$m counts with the SCUBA-2 450 $\mu$m counts from the literature obtained by blind extraction \citep{2024Gao, zavala2017, Wang2017, Chen2013b, Casey2013, Geach2013}. All 450 $\mu$m counts have been corrected according to the correction factors indicated by the respective papers if necessary as well as by a final correction factor from \citet{2024Gao}. In addition, we applied a flux correction factor of 0.84 to convert the SCUBA-2 450 $\mu$m flux to the SPIRE 500 $\mu$m flux. Given the much higher SCUBA-2 resolution of 7.9$\arcsec$ (compared to the SPIRE resolution of 36.6$\arcsec$), the 450 $\mu$m blind counts for bright fluxes are much more reliable. We find an excellent agreement between our number counts from the super-resolved sources and the 
500 $\mu$m counts converted from the SCUBA-2 counts down to almost the $1\sigma$ confusion noise level, particularly with the latest and the most reliable counts from \citet{2024Gao}.

%--------------------------------------------------------------------
\section{Summary and Conclusions}\label{sect:Conclusions}
%----------------
% Summary
%----------------
In this paper, we presented a DL-based SR method based on a Transformer network following an auto-encoder structure to deblend and denoise \textit{Herschel}/SPIRE 500 $\mu$m observations. We used a combined loss function with the goal of reproducing the overall image characteristics and source fluxes. This loss function was designed to be similar to what \citet{Lauritsen_SR_2021} used but with the aperture loss component being differentiable within the machine-learning frameworks. Subsequently, the network was trained on realistic mock data obtained from the SIDES and SHARK simulations. The input data, consisting of the \textit{Spitzer}/MIPS 24 $\mu$m and \textit{Herschel}/SPIRE 250, 350, 500 $\mu$m, have been generated with realistic PSFs and instrumental noise mimicking observations as much as possible. We evaluate the performance of our model on a simulated test set as well as real \textit{JCMT}/SCUBA-2 450 $\mu$m observations in the COSMOS field. Our main results are summarised below:

%----------------
% Conclusions & Strengths
%----------------
\begin{enumerate}
    \renewcommand{\labelenumi}{\it \roman{enumi})}
    \item Validation using the simulations demonstrates that our  method can super-resolve the SPIRE 500 $\mu$m map with high flux accuracy, showing only a mean 5\% underestimation for sources brighter than $\sim8$ mJy, and high positional accuracy, with most sources recovered within 1$\arcsec$ of the target sources. The application of SR yields significantly better results compared to blind source extraction methods on the native SPIRE 500 $\mu$m map.
    
    \item Our method is highly reliable, with a cumulative reliability of over 90\% for extracted super-resolved sources above $\sim3$ mJy, based on the simulation test set. Additionally, for all target sources with fluxes $\gtrsim$ 5 mJy, we achieve a high completeness of $\gtrsim$90\%.

    \item We super-resolved the SPIRE 500 $\mu$m observations of the COSMOS field and compared our results with the the SCUBA-2 450 $\mu$m observations (which have a resolution of 7.9$\arcsec$). We find very good agreement with the SCUBA-2 measured fluxes. In addition, the corrected number counts from our super-resolved sources are consistent with the SCUBA-2 counts, demonstrating again the high completeness and reliability of our method. 

    \item The inference time of our method, $\sim1$ second per deg$^2$, is orders of magnitudes faster than traditional deblending methods, allowing for fast application on survey on the scale of ten to hundreds of deg$^2$. Moreover, our method is easy to use, requiring no selection of prior sources. 

\end{enumerate}
%----------------
% Future Outlook
%----------------
Our methodology can be easily extended with new wavelength observations with almost no impact on the inference time. In the next phase, we will expand the super-resolved maps to all SPIRE bands, i.e. super-resolving the 250, 350, 500 $\mu$m bands simultaneously. We will also apply this on a large field observed by both \textit{Herschel} and \textit{Spitzer} and compare with high spatial resolution radio observations from LOFAR, utilising the FIR-radio correlation. The SPIRE instrument has observed around 380 deg$^2$ at varying depths \citep{2014MNRAS.444.2870W}. Super-resolved catalogues with their improved positional and photometric accuracy would provide a significant improvement in the number of 250, 350 and 500 $\mu$m galaxies that can be cross-matched with other catalogues yielding a more complete multi-wavelength view that can provide more accurate SEDs and hence physical properties of the galaxies. This is especially helpful for high-$z$ DSFGs where they appear the brightest in the SPIRE bands. To further improve our methodology, it could be worthwhile investigating the feasibility of a loss component that can lead to the inclusion of the observational images in training or as a fine-tuning step.  
For our use case, the loss function could consist of a neural network that convolves the super-resolved images back to the native resolution and compare the result. Deviations due to the instrumental noise should sum to zero and its impact should be minimal. 
Finally, future work could extend the input image set by incorporating other observations such as those from \textit{Euclid}. 
%--------------------------------------------------------------------
\newpage
\begin{acknowledgements}

We made use of an existing Keras Vision Transformer implementation available at \url{https://github.com/yingkaisha/keras-vision-transformer/tree/main}.
      This publication is part of the project ‘Clash of the titans:
deciphering the enigmatic role of cosmic collisions’ (with project number VI.Vidi.193.113 of the research programme Vidi which is (partly) financed by the Dutch Research Council (NWO).

Herschel is an ESA space observatory with science instruments provided by European-led Principal Investigator consortia and with important participation from NASA.

We thank the Center for Information Technology of the University of Groningen for their support and for providing access to the Hábrók high performance computing cluster.

We thank SURF (www.surf.nl) for the support in using the National Supercomputer Snellius.

\end{acknowledgements}

\bibliography{references}

\begin{appendix}
\section{Description of training}
\begin{figure}[ht] 
    \centering
    \includegraphics[width=\linewidth]{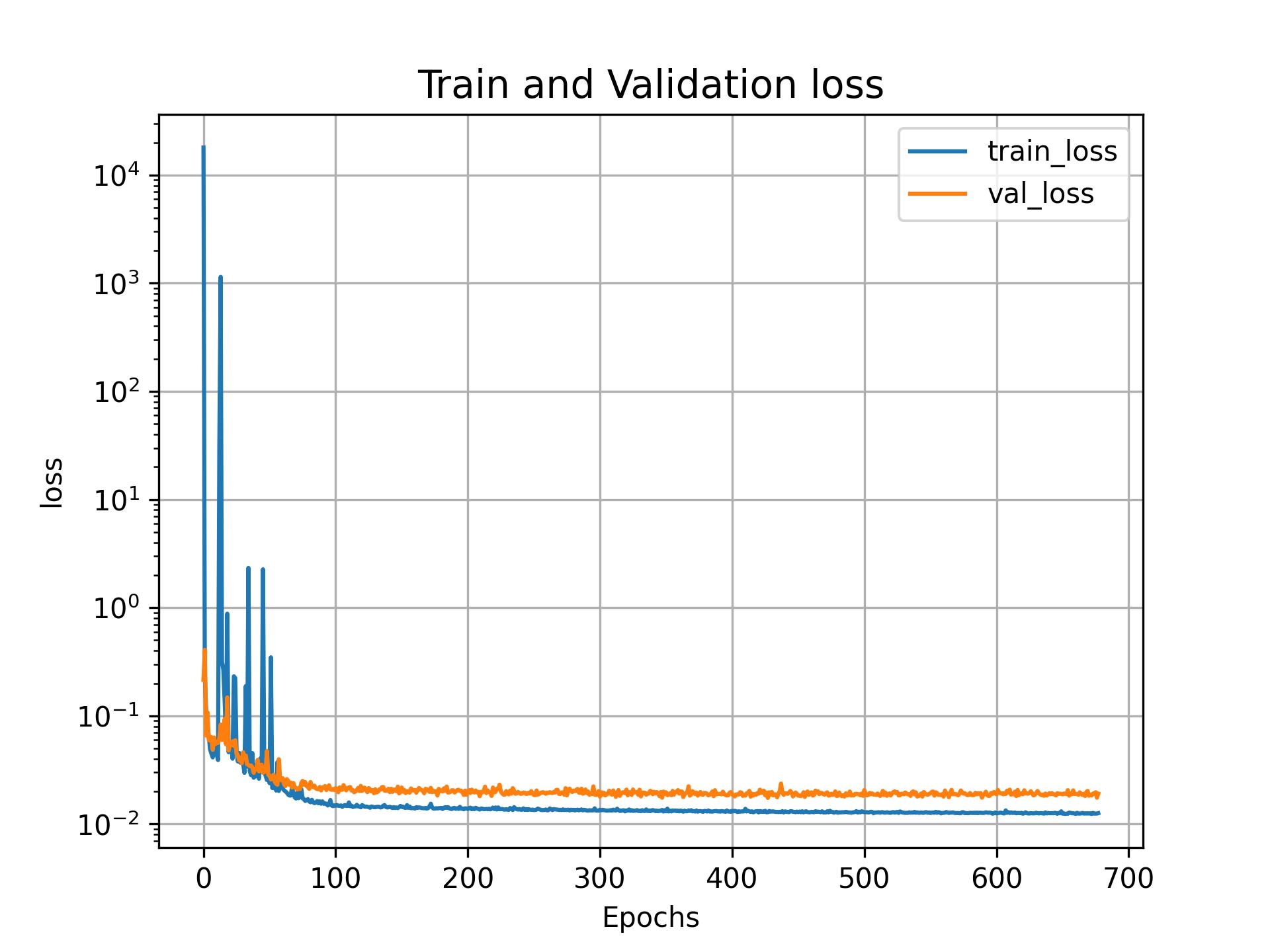} 
    \caption{The training history showing the training and validation loss as a function of epoch number. During each epoch the model sees the training data only once and the samples are randomly augmented. Some high spikes are shown which are suppressed by the implementation of gradient clipping.}
    \label{fig:trainingcurves}
\end{figure}

The training curves of our main model results are shown in Fig. \ref{fig:trainingcurves}. These curves show that early stopping has triggered in time as overfitting is starting to show. Moreover, gradient clipping is successful in suppressing gradient spikes.
\section{Performance based on Spitzer MIPS 24$\mu$m flux}
\label{sec:Performance based on Spitzer MIPS}
\begin{figure}[ht]
    \centering
    \includegraphics[width=0.95\columnwidth]{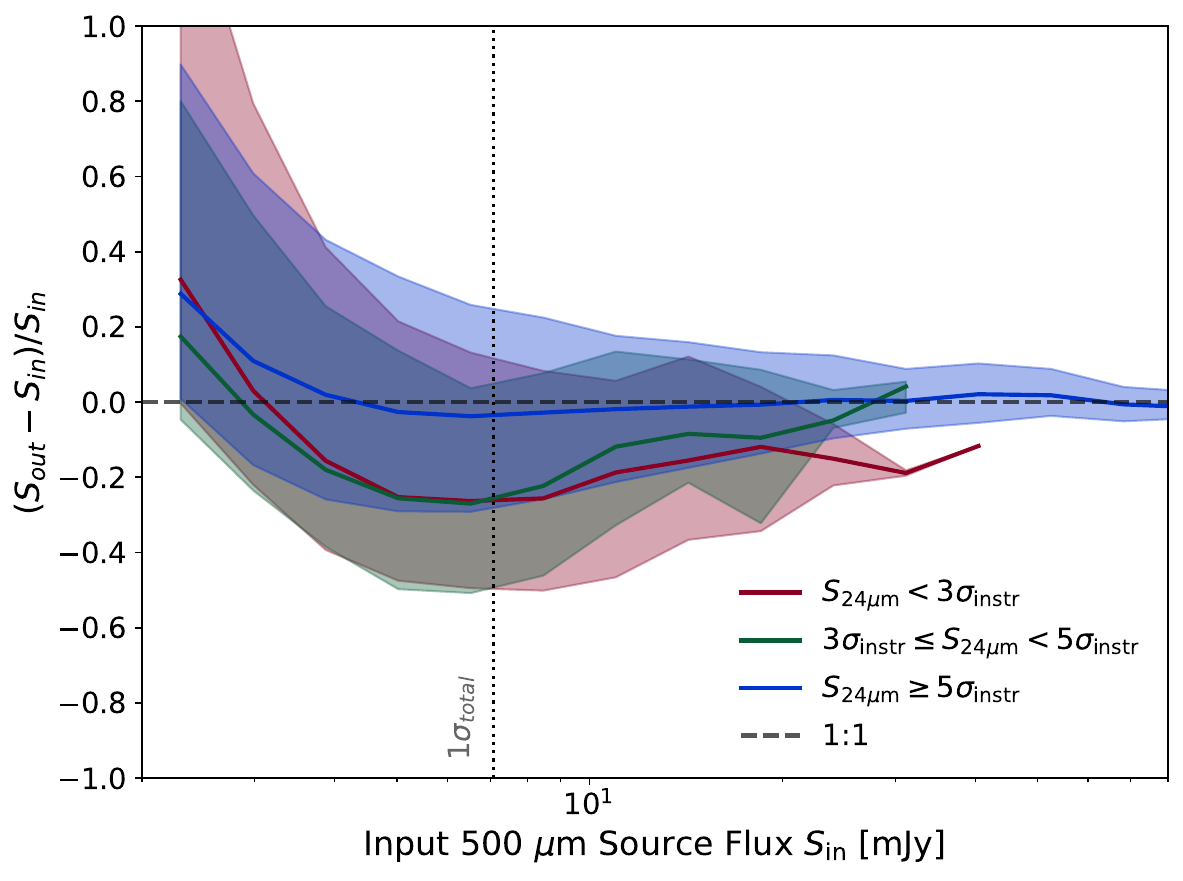} 
    \caption{Fractional flux difference as a function of the input flux $S_{in}$ decomposed for varying levels of Spitzer MIPS 24$\mu$m flux. The measured flux $S_{out}$ corresponds to the super-resolved simulated source fluxes. The 50th (solid line), 16th and 84th (shaded regions) percentiles are shown for three Spitzer MIPS 24$\mu$m flux intervals as a function of the simulated instrument noise level. The dashed horizontal line indicates a 1:1 agreement.}
    \label{fig:InputFluxReproduction_mips_decomposed}
\end{figure}

In Section \ref{sec:FIR/sub-mm observations}, we mention that $\sim$80\% of the 24 $\mu$m emission can be attributed to $z\leq$2 sources. To investigate whether there is degraded performance for fainter 24 $\mu$m sources, we investigate the flux reproduction performance as a function of 24 $\mu$m flux. In Fig. \ref{fig:InputFluxReproduction_mips_decomposed}, we show the decomposition of our main super-resolution results from the simulated test data (Fig. \ref{fig:InputFluxReproduction}) as a function of the Spitzer MIPS 24$\mu$m instrument noise. We show the flux reproduction for sources with $S_{24\mu m} < 3\sigma_{instr}$, $3\sigma_{instr} \leq S_{24\mu m} < 5\sigma_{instr}$ and $S_{24\mu m} \geq 5\sigma_{instr}$. We find that the 500 $\mu m$ fluxes for sources with $S_{24\mu m} \geq 5\sigma_{instr}$ can be predicted extremely well up to 1$\sigma_{total}$ with a slight underestimation of a few percent. Moreover, we find indeed a worse flux reproduction for sources $S_{24\mu m}  < 5\sigma_{instr}$, which shows that our methodology may underestimate 500 $\mu m$ fluxes increasingly more for high-z sources. Nevertheless, we note that high-z sources with $S_{24\mu m} \geq 5\sigma_{instr}$ could still be reproduced well. 
\section{Dependence on Noise}
\begin{figure}[ht]
    \centering
    \includegraphics[width=0.95\columnwidth]{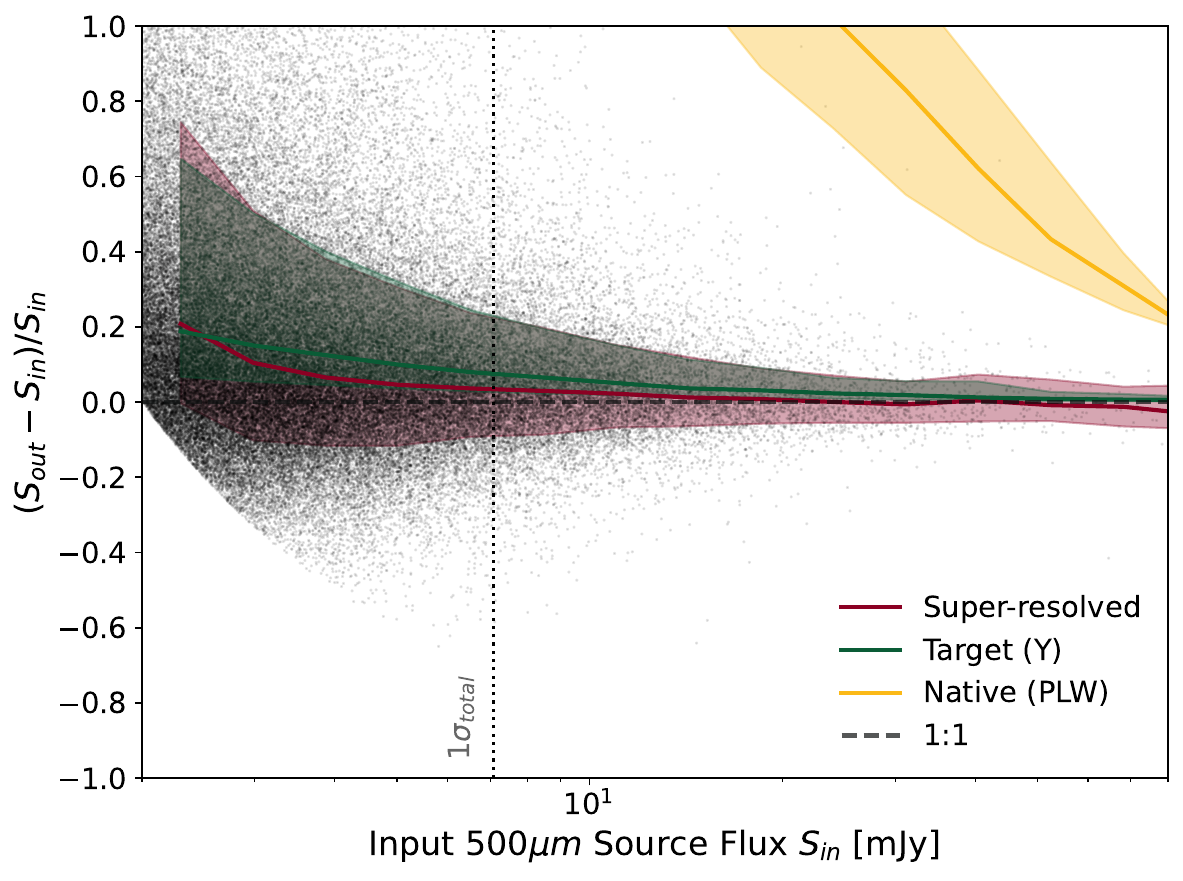} 
    \caption{Fractional flux difference as a function of the input flux $S_{in}$. The measured flux $S_{out}$ corresponds to the native, target and super-resolved source fluxes. The 50th (solid line), 16th and 84th (shaded regions) percentiles are shown for each output catalogue. The dashed horizontal line indicates a 1:1 agreement. The grey points correspond to the super-resolved sources. The super-resolved images are predicted from instrument noise-free input images. Here, the vertical $1 \sigma$ line represents the total noise of real 500$\mu$m observations and that used in our main results. }
    \label{fig:InputFluxReproductionSmoothed}
\end{figure}

We trained on a 120 deg$^2$ dataset without instrument noise in the input images to investigate the slight flux underestimation and to see how well we can reproduce the target in a perfect setting with only confusion noise. This also serves as a test of the capability of only using 4 broad-band flux measurements to predict the 500 $\mu$m fluxes at the target resolution of 7.9$\arcsec$. In Fig. \ref{fig:InputFluxReproductionSmoothed}, we show the results on the corresponding noise-free test set. We find that the slight underestimation previously seen with instrument noise in the input images (Fig. \ref{fig:InputFluxReproduction}) vanishes which clearly shows that instrument noise is able to affect training and deteriorate flux reproduction near and below the total noise $1\sigma$. Moreover, we find that our model is able to approximately reproduce the expected flux boosting at the target resolution as well as the upper statistical uncertainties.

\begin{figure}[ht] 
    \centering
    \includegraphics[width=\linewidth]{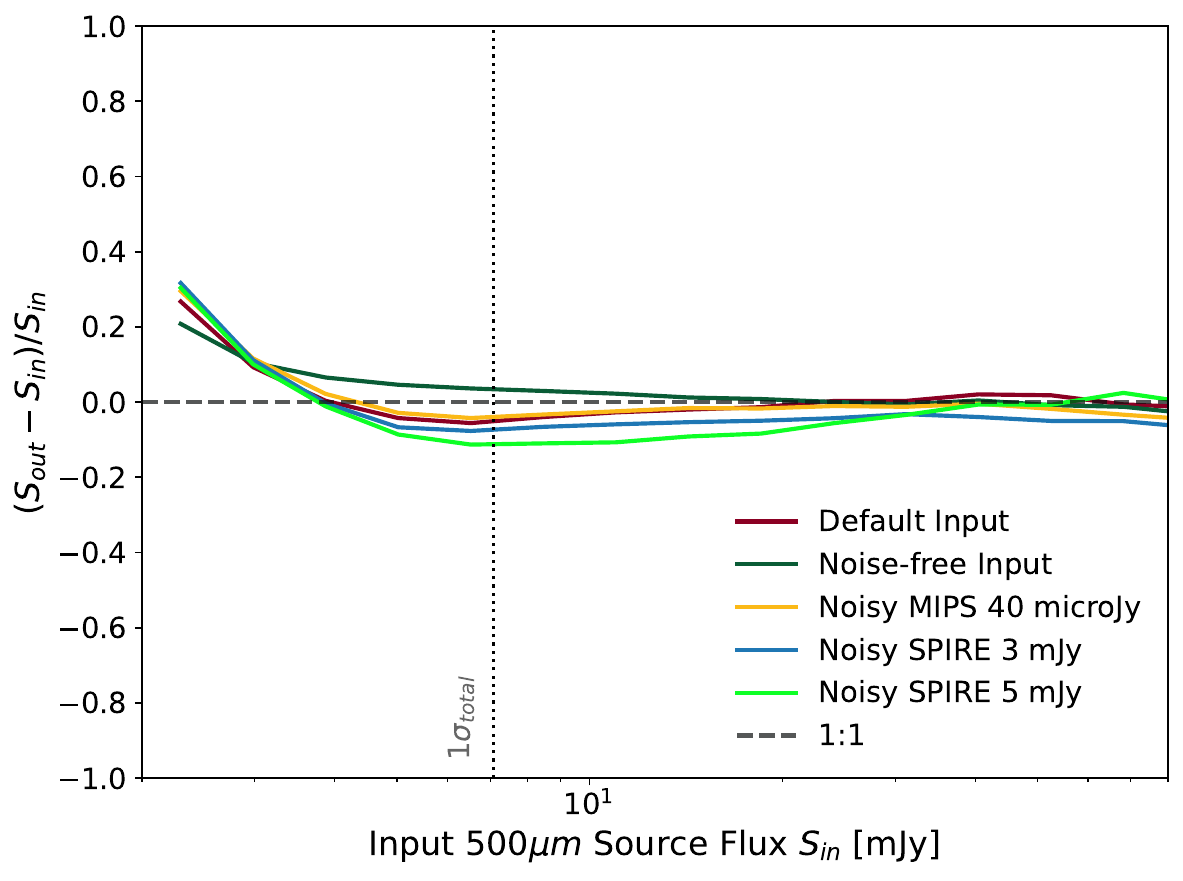} 
    \caption{Fractional flux difference as a function of the input flux $S_{in}$. The measured flux $S_{out}$ corresponds to the super-resolved source fluxes. The coloured lines represent the 50th percentile for each output catalogue. Each line depicts the performance on a test set with a given instrument noise level. The legend only highlights the new noise level of the specific band(s) with respect to the default configuration described by Table \ref{table:characteristics_combined} and represented by the red curve. The dashed horizontal line indicates a 1:1 agreement.
    }
    \label{fig:InputFluxReproduction_varying_noise}
\end{figure}

\begin{figure*}[ht]
  \centering\includegraphics[width=18cm]{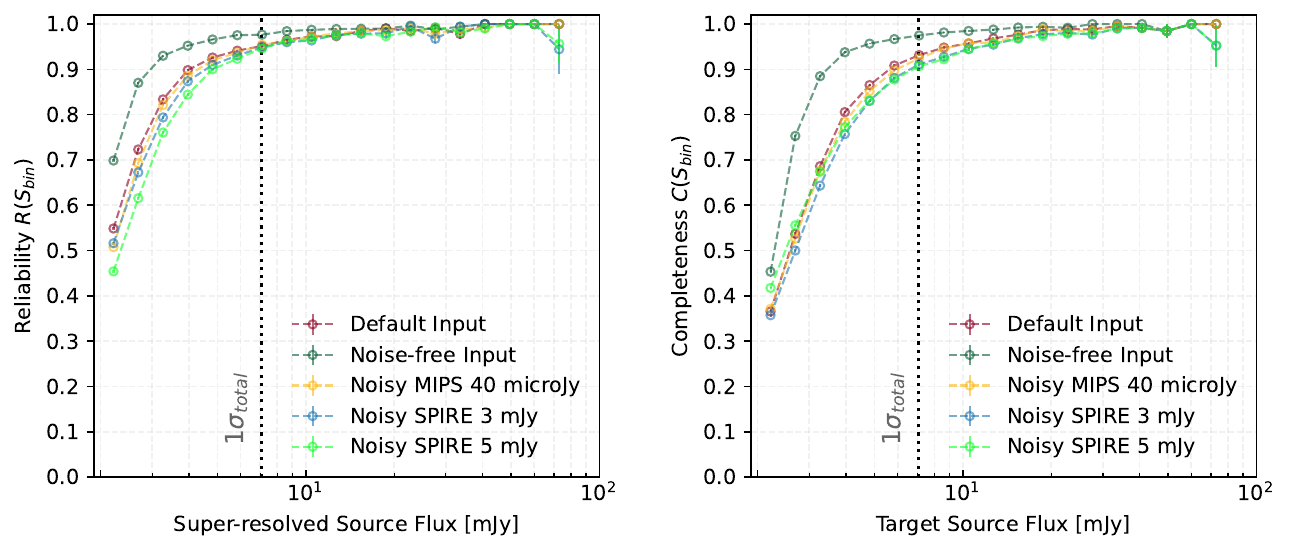}
  \caption{Left: Reliability of the super-resolved sources as a function of the detected source flux (Only R($S_{bin}$) is shown). Right: Completeness of the target sources as a function of the target source flux (Only C($S_{bin}$) is shown).
  Error bars are the $1\sigma$ uncertainties corresponding to a binomial distribution. }
  \label{fig:sim_CR_varying_noise}
\end{figure*}

Moreover, to investigate the impact of varying levels of instrument noise, we trained on several 120 deg$^2$ datasets with varying noise levels. For two datasets we increased the $1\sigma_{instr}$ of the three SPIRE bands to 3 mJy and 5 mJy with that of Spitzer/MIPS 24$\mu$m remaining unchanged as described by Table \ref{table:characteristics_combined}. For another dataset, we increased the $1\sigma_{instr}$ noise of the Spitzer/MIPS 24$\mu$m input images to 40$\mu$Jy with the SPIRE bands fixed at their default noise levels as described by the Table. In Fig. \ref{fig:InputFluxReproduction_varying_noise}, we demonstrate the flux reproduction compared to the input test catalogue. Overall, we find that our methodology is robust against varying noise levels that can be expected in most real-world observations \citep{2008AJ....135.1050S, 2014MNRAS.444.2870W}. Increasing the Spitzer/MIPS 24$\mu$m instrument noise has a negligible effect on the flux reproduction while increasing the Herschel/SPIRE instrument noise slightly decreases the flux reproduction performance by 5-10\% for fluxes below 20 mJy whilst there is negligible impact on the bright fluxes ($\geq 30$ mJy) where the differences are mostly from low number counts. This behaviour at fainter fluxes from higher input noise in the SPIRE bands is expected as these fluxes are most important in predicting the fluxes at 500 $\mu$m in the super-resolved images. In Fig. \ref{fig:sim_CR_varying_noise}, we show the impact on the reliability and completeness. Here, we find a small trend, with both the reliability and completeness decreasing with increasing instrument noise levels in the input bands. A bit unexpectedly, increasing the instrument noise for Spitzer/MIPS 24$\mu$m by more than a factor 2 has a much lower impact than increasing the noise in the SPIRE bands by 1 or 2 mJy. It is likely that 40 $\mu$Jy is still deep enough for enough sources to be detected and recovered and we expect that increasing it further will start to have a detrimental impact on the reliability and completeness. Finally, training on a hybrid dataset containing varying noise levels in the input images could make our method more robust against varying noise levels in real-world observations.

\section{Performance on Extended Galaxies}
\label{sec:Performance on Extended Galaxies}
\begin{figure*}[ht]
  \centering\includegraphics[width=16cm]{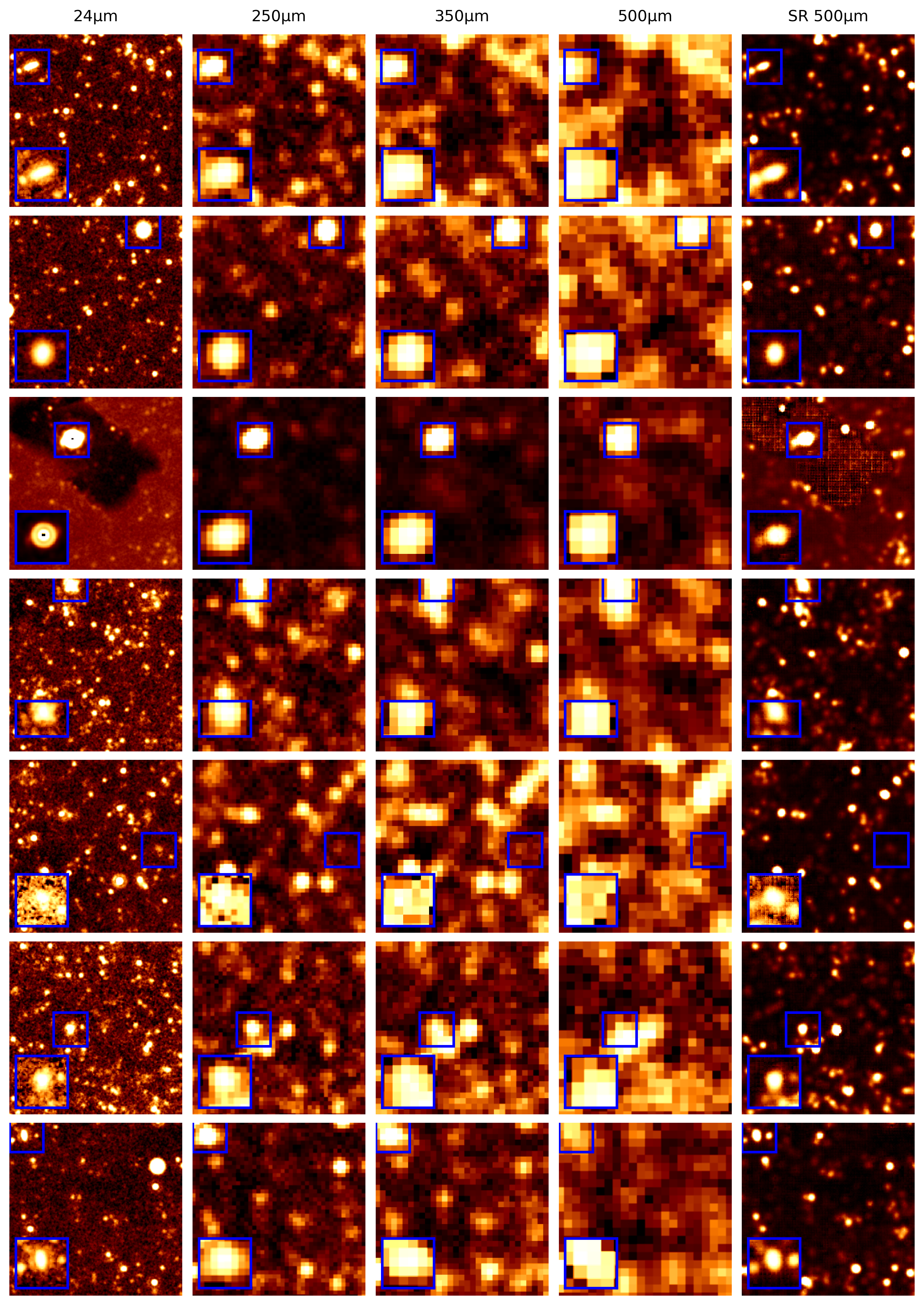}
  \caption{SR performance on observed images in the COSMOS field containing extended sources as listed in XSC. Across the rows we show the COSMOS images with each containing one extended source per XSC. First column: The native Spitzer/MIPS 24 $\mu$m images. Second-Fourth columns: The native Herschel/SPIRE 250, 350 and 500 $\mu m$ images. Fifth column: The super-resolved Herschel/SPIRE 500 $\mu m$ images. The blue boxes highlight regions centered on the extended sources. We used Arcsinh contrast stretching to make fainter structures more visible.}
  \label{fig:extended_source_analysis}
\end{figure*}

In this section, we briefly demonstrate the performance of our model on extended sources. We used the 2MASS All-Sky Extended Source Catalog (XSC) \citep{2MASS_ext} to obtain a selection of 7 extended galaxies that would be atleast extended in the Spitzer/MIPS 24$\mu$m band in the COSMOS field. We queried the catalogue to only include sources covered by the Spitzer/MIPS footprint shown in Fig. \ref{fig:cosmos_contour_projection} and those that have a 20 mag isophotal fiducial elliptical apparent semi-major axis greater than 9 arcseconds (column r\_k20fe). In Fig. \ref{fig:extended_source_analysis}, we illustrate the cutouts across the four input bands as well as the super-resolved cutout. Here, we used an asinc stretch in the highlighted regions to magnify the morphology. Visually, the model seems to be able to cope well with extended sources despite expectations with the model not having seen extended morphologies during training. Nevertheless, with the current input setup, we can not conclude whether the SR method is able to reproduce extended source morphologies and fluxes on a similar level as we demonstrated with point-sources. This is mainly due to faint features and no knowledge on the morphology. The angular resolution of Spitzer/MIPS is likely too low such that the galaxies look like slightly elongated point sources with no features visible.

\end{appendix}
%--------------------------------------------------------------------

\end{document}